\documentclass[12,preprint]{aastex}

\newcommand{\eight}{850~$\mu$m}%
\newcommand{\four}{450~$\mu$m}%
\newcommand{\clfind}{\emph{clfind2d}}%

\begin{document}

\title{High Mass Star Formation II: The Mass Function of Submillimeter Clumps in M17}

\author{Michael A. Reid and Christine D. Wilson}
\affil{Department of Physics and Astronomy, McMaster University, 
Hamilton, ON, L8S 4M1, Canada \& Harvard-Smithsonian Submillimeter Array 
Project, 645 N A'ohoku Pl., Hilo, HI, 96720, USA}

\begin{abstract} 

	We have mapped a $\sim 5.5\times5.5$~pc portion of the M17 massive 
star-forming region in both 850 and \four\ dust continuum emission using 
the Submillimeter Common-User Bolometer Array (SCUBA) on the James Clerk 
Maxwell Telescope (JCMT).  The maps reveal more than 100 dusty clumps with 
deconvolved linear sizes of $\sim$0.05--0.2~pc and masses of 
$\sim$0.8--$120~M_{\odot}$, most of which are not associated with known 
mid-infrared point sources.  Fitting the clump mass function with a double 
power law gives a mean power law exponent of $\alpha_{\rm high} = 
-2.4\pm0.3$ for the high-mass power law, consistent with the exponent of 
the Salpeter stellar mass function.  We show that a lognormal clump mass 
distribution with a peak at $\sim$4$~M_{\odot}$ produces as good a fit to 
the clump mass function as does a double power law.  This 4$~M_{\odot}$ 
peak mass is well above the peak masses of both the stellar initial mass 
function and the mass function of clumps in low-mass star-forming regions.  
Despite the difference in intrinsic mass scale, the \emph{shape} of the 
M17 clump mass function appears to be consistent with the shape of the 
core mass function in low-mass star-forming regions.  Thus, we suggest 
that the clump mass function in high-mass star-forming regions may be a 
scaled-up version of that in low-mass regions, instead of its extension to 
higher masses.

\end{abstract}


\section{INTRODUCTION}

	Numerous unresolved aspects of the theory of star formation depend 
on a detailed knowledge of the structure of the interstellar medium in 
star-forming regions.  For example, it is well established that the 
turbulent fragmentation of molecular clouds gives rise to a variety of 
small-scale structures, some of which collapse gravitationally to form 
stars (e.g., \citealt{ts98,m01,klessen01,kb00}).  However, it is not clear 
what determines the final, relatively invariant distribution of stellar 
masses.  In order to identify the processes which set the stellar mass 
function, we must first know how the masses of pre-stellar cores and 
clumps are distributed.  

	Submillimeter continuum observations of low-mass star-forming 
regions have suggested that the mass function of 
\mbox{$10^{3}$--$10^{4}$~AU-scale} dense clumps with masses less than a 
few $M_{\odot}$ resembles the stellar initial mass function (IMF) over the 
same mass range \citep{ts98,dj2000,m01}.  It is less clear how the mass 
function of massive clumps/cores relates to that of massive stars, mostly 
due to a lack of submillimeter continuum observations of massive 
star-forming regions at sufficiently high resolution.  The relationship 
between the masses of cores and stars is especially important at higher 
masses because it could help to distinguish between competing theories 
of high-mass star-formation.  According to the two prevailing theories, 
massive stars may form either by the monolithic collapse of individual 
molecular cloud cores \citep{mt02,mt03,krum05}, aided by the formation of 
a disk and an optically thin outflow cavity, or by the competitive 
accretion of material from the larger reservoir of an entire proto-cluster 
\citep{bon97,bon01a,bon01b,bb02}.  In principle, we should expect a closer 
resemblance between the mass functions of cores and stars in the first 
case than in the second.  Whatever the formation mechanism of massive 
stars, it is likely to have signature in the clump/core mass function.

	To date, only a handful of studies have measured the submillimeter 
continuum mass function of cores and clumps at and beyond the high-mass 
end of the stellar mass range (\citealt{kerton,tot,msl03,paperi}, 
``Paper~I'' hereafter). Because the mean distance to massive star-forming 
regions is large compared to their low-mass counterparts, most studies of 
such regions have been sensitive only to spatial and mass scales 
significantly larger than those of pre-stellar clumps 
\citep{msl03,mookerjea}.  In this paper and Paper~I, we present 
submillimeter continuum maps of two relatively nearby massive star-forming 
regions, in which it is possible to discern structures which may form 
individual stars.  Paper~I concentrated on NGC~7538, a massive 
star-forming region at a distance of 2.8~kpc.  The focus of this paper is 
M17 which, at a distance of $1.6\pm0.3$~kpc \citep{n01}, is near enough to 
permit us to discern candidate pre-stellar condensations.  The 
$\sim$8\arcsec~beam of the James Clerk Maxwell Telescope at 
\four\ translates to a linear resolution of $\sim$0.06~pc 
($\sim$1.3$\times10^{4}$ AU) at 1.6~kpc, which is comparable to the 
resolution obtained in similar observations of nearby low-mass 
star-forming regions (typically 0.01--0.03~pc).  

	A large-scale CO $J$=1$\rightarrow$0 map of M17 shows it to be 
divided into northern and southern parts whose combined mass exceeds 
$3\times10^{4} M_{\odot}$ \citep{l76}.  The best-known features of the 
region are the well-studied photon-dominated region and the molecular 
cloud it borders, M17SW (e.g., \citealt{bt01,sg90,fmc84}).  M17SW exhibits 
many signs of massive star formation, including an ultracompact {\sc Hii} 
(UC{\sc Hii}) region, numerous locations of maser activity, and a cluster 
of young O and B type stars \citep{hanson97,chini2000l,bt01}.  In their 
survey of 69 stars in M17, \citet{cw98} found 20 stars with strong 
infrared excesses and luminosities more than 60 times those of Class~I 
sources in $\rho$~Oph.  They identify these sources as candidate massive 
Class~I analogs, which suggests that there is a significant population of 
young, early type massive stars in the region.

	Because we are interested in studying the earliest stages of
massive star formation, rather than study M17SW, where massive star
formation is already well underway, we focus instead on the northern part
of M17, in the region around the compact source M17N.  The CO maps of
\citet{l76} and \citet{wilson99} show the M17N region to be rich in
molecular gas, but radio continuum studies (e.g.,  \citealt{lada76}) show
it to be relatively devoid of free-free emission, which suggests it is
relatively free of ionizing massive stars.  Most of the detected radio
continuum emission originates in the UC{\sc Hii} region associated with
M17N \citep{wilson79}, which coincides with an H$_{2}$O maser
\citep{jgd81}.  At least 5 infrared point sources in the immediate
vicinity of M17N are believed to be embedded stars
\citep{klein99,henning98}.

	\citet{klein99} made a 3\arcmin$\times$4\arcmin~map of the region
around M17N at 1.3~mm with the IRAM~30m telescope, which revealed the
extended, clumpy nature of the dust continuum emission.  In order to
identify new cold, dusty clumps in the M17N region and to characterize
their mass function, we have obtained larger, more sensitive maps of the
region at \four\ and \eight.  In \S \ref{sec:obs}, we describe our
observations and data reduction techniques.  In \S \ref{sec:props},
we discuss the properties of the clumps.  Section \ref{sec:m17massfunc} is
devoted to a detailed analysis of the M17 clump mass function, including
comparisons to the mass function of CO clumps and the stellar IMF.  In
Paper~III, the results of Papers~I and II will be compared quantitatively
to the clump mass functions measured in other star-forming regions, with
the goal of determining a general functional form for the clump mass
function.

\section{OBSERVATIONS AND DATA REDUCTION}
\label{sec:obs}

We used the Submillimeter Common-User Bolometer Array on the James Clerk 
Maxwell Telescope to map an approximately 
12\arcmin$\times$12\arcmin~(5.5$\times$5.5 pc) region of M17.  The data 
were acquired on the nights of 2003 March 17 and April 16 with a total 
on-source integration time of about 5 hr.  The data were taken in the 
standard scan-mapping mode with chop throws of 30\arcsec, 44\arcsec, and 
68\arcsec~in both right ascension and declination.  Pointing checks were 
performed once per hour and sky dips once every hour or two.  The rms 
pointing accuracy was approximately 1.7\arcsec.  Mars was used as the flux 
calibrator and was observed every 3--4 hours; the derived uncertainties in 
the gain calibrations were 4\% at \four\ and 12\% at \eight.  The sky 
opacities at \eight\ and \four\ were calculated using polynomial fits to 
the combination of the JCMT skydips and the 225 GHz zenith optical depth 
measurements from the Caltech Submillimeter Observatory.  The mean 225~GHz 
zenith optical depths on the two nights were 0.093 and 0.054, 
respectively.  The mean residuals from the polynomial fits to the optical 
depth data translate into a systematic uncertainty in the fluxes of less 
than 1\%.

	We used the standard techniques for reducing and calibrating SCUBA 
scan maps (detailed in Paper~I), including map reconstruction with the 
Emerson2 algorithm \citep{emerson2}.  The mean rms flux measured in 
emission-free regions of the maps is 0.027 Jy~beam$^{-1}$ at \eight\ and 
0.32 Jy~beam$^{-1}$ at \four.  These rms fluxes are somewhat higher than 
those from Paper~I (0.021~Jy~beam$^{-1}$ and 0.18~Jy~beam$^{-1}$ at \eight\ 
and \four, respectively), but M17 is nearly twice as close as NGC~7538, 
making the M17 maps significantly more sensitive in terms of luminosity.  
The half power beam widths at the two wavelengths were 15.4\arcsec~and 
8.5\arcsec, respectively.  We used the ``image flattening'' technique 
described in Paper~I to remove spurious large-scale structure introduced 
during reconstruction of the scan map using the Emerson2 technique.  The 
flattening technique consists of clipping the brightest emission peaks 
from the map, smoothing the remainder to approximate the large-scale 
background of the map, and subtracting the result from the original map.  
The technique preserves all structure on scales smaller than approximately 
twice the size of the largest chop throw, or about 2\arcmin.  Because the 
strongest source in the maps (M17N) is considerably weaker than the 
strongest source in the NGC~7538 region, the small artifacts introduced by 
the flattening technique are even smaller in the M17 maps than in the 
NGC~7538 maps.  The only visible effect of the flattening procedure is 
a slight reduction in the brightness of the structure extending northward 
from M17N.  The \eight\ and \four\ maps are shown in 
Figures~\ref{fig:m17850map} and \ref{fig:m17450map}, respectively.

\section{PROPERTIES OF THE M17 CLUMPS}
\label{sec:props}

\subsection{Clump Identification and Contamination Corrections}

	We used \clfind\ \citep{clfind} to locate the clumps in our maps 
and determine their boundaries and integrated fluxes.  To be considered a 
clump, an object must satisfy three criteria: its area must be at least 
half that of the beam, its peak flux must be at least 5~$\sigma$, and it 
must have a flux of $> 3~\sigma$ in every pixel.  Using \clfind\ with 
these detectability criteria, we identified 121 clumps in the \eight\ image 
and 101 in the \four\ image.  The properties of the \eight\ and 
\four\ clumps 
are listed in Tables \ref{tab:m17850clumps} and \ref{tab:m17450clumps}, 
respectively.  Note that the clump radii stated throughout this paper are 
the effective radii of the clumps, deconvolved from the beam.  The 
effective radius of a clump is defined as the radius of a circle with an 
area equal to the projected area of the clump.

	As we are principally interested in the dust continuum emission of
each clump, we must correct the fluxes for three likely sources of
contamination: radio continuum emission, CO $J$=3$\rightarrow$2 line
emission in the \eight\ filter, and the sometimes substantial JCMT error
beam.

\subsubsection{Radio Continuum Corrections to S$_{\rm int}$}

	Radio continuum contamination is principally a concern in M17N, 
which harbors an UC{\sc Hii} region \citep{wilson79}, and in the 
southeastern part of the map, where some of the dust continuum emission is 
coincident with the ``Northern Bar'' structure seen in numerous radio 
continuum studies (e.g., \citealt{bt01,fmc84}).  Combining the measured 
1.4 and 3.5~cm radio continuum fluxes of M17N from \citet{lada76} and 
\citet{wilson79}, respectively, with the assumption that the radio 
continuum emission follows a power law of the form $S_{\nu} \propto 
\nu^{-0.1}$, we estimate the radio continuum contamination of M17N at 
\eight\ to be $\sim$30\%, or $\sim$2.2~Jy.  We used the 6~cm map of 
\cite{fmc84} to estimate radio continuum corrections to the \eight\ fluxes 
of the 18 clumps which are coincident with the Northern Bar, again 
assuming the $S_{\nu} \propto \nu^{-0.1}$ scaling relationship.  The 
corrections to the \eight\ fluxes derived in this way ranged from 4 to 
45\%, but were typically less than 20\%.  Radio continuum corrections to 
the \four\ fluxes were found to be negligible.  The remainder of the 
clumps in the map are not coincident with any known radio continuum 
sources (see, for example, the map of \citealt{lada76}).

\subsubsection{CO~$J$=3$\rightarrow$2 Corrections to S$_{\rm int}$}

	The derivation of CO~$J$=3$\rightarrow$2 contamination corrections 
is complicated by the lack of a CO~$J$=3$\rightarrow$2 map with the same 
spatial resolution and coverage as our continuum maps.  The closest match 
is the CO~$J$=3$\rightarrow$2~map of \citet{wilson99}, which completely 
covers our map, but is not fully sampled.  The \citet{wilson99} map was 
made using the position-switching technique instead of chopping.  Thus, it 
traces the total column of CO~$J$=3$\rightarrow$2, while our chopped maps 
can suffer contamination only from CO~$J$=3$\rightarrow$2~emission on 
scales smaller than about twice our largest chop throw, or $\sim$2\arcmin.  
Comparing our maps to those of \citet{wilson99} reveals that most of the 
CO structure in the region of overlap occurs on large spatial scales, 
greater than 2\arcmin.  Hence, most of this emission would be chopped out 
in our map, and would not contaminate the \eight\ fluxes.  We expect the CO 
contamination to be worst close to M17N, where the CO emission peaks.  We 
estimate that the CO contamination of the \eight\ flux of the clump 
associated with M17N would be about 26\% if we were sensitive to the total 
column of CO, but half or less than that in the chopped map.  Because the 
degree of CO contamination is difficult to estimate, but likely less than 
10\% in most clumps, we have not applied CO corrections to our data.

\subsubsection{Error Beam Corrections to S$_{\rm int}$}

	The JCMT beam can be reasonably approximated by a sum of Gaussian
components: a strong, narrow primary beam and a broad, faint error beam.  
In Paper~I, we described our method for fitting the beam structure using 
our calibrator observations.  In this work, we used the same beam
fits as in Paper~I to correct the integrated flux of each clump.  The
error beam corrections at \eight\ ranged from 3 to 14\% with a mean of
10\%.  The equivalent corrections at \four\ ranged from 3 to 18\% with a
mean value of 11\%.

\subsection{Clump Spectral Indices and Temperatures}
\label{sec:alphaandt}

	We have used the combination of our \eight\ and \four\ maps to 
measure the spectral index, $\alpha$, of each clump, and thereby to 
estimate its temperature.  The spectral index in a given pixel of the map 
is 
defined as:

\begin{equation}
\label{eq:specindex}
\alpha = \frac{\log(S_{\rm 450 \mu {\rm m}}/S_{\rm 850 \mu {\rm m}})}{\log(850/450)}~~.
\end{equation}

\noindent As described in detail in Paper~I, it is necessary first to 
match the beams of the two images by convolving each map with the beam at 
the other wavelength.  This method not only ensures that the resolutions 
of the two maps match, but also that their error beam structures match.  
We applied the \clfind\ boundaries of the \eight\ clumps to the spectral 
index map to calculate the flux-weighted mean spectral index of each 
\eight\ clump.  In computing the flux-weighted mean spectral index, we 
weight the spectral index of each pixel by the flux of the corresponding 
pixel in the convolved (i.e. resolution-matched) \eight\ map.  Only pixels 
detected at $> 3 \sigma$ at both \eight\ and \four\ are included in the 
mean.  The mean spectral indices so calculated are given in Table 
\ref{tab:m17850clumps}.  The $\sim$18\arcsec~beam of the spectral index 
map prohibits direct comparison between it and the \four\ map, whose beam 
is less than half as broad.  The broader $\sim$15\arcsec~beam of the 
\eight\ map permits direct comparisons with the spectral index map.  
Making this comparison, we find, as in Paper~I, that continuum emission 
peaks in the \eight\ map are strongly correlated with peaks in the spectral 
index map.

	We can use the derived spectral indices to estimate the mean
temperature of each clump.  The mean flux ratio of a clump is related
to its mean dust temperature, $\langle\mbox{T}_{\rm dust}\rangle$, and
mean dust emissivity index, $\langle\beta\rangle$ by:

\begin{equation}
\label{eq:tdust}
\left<\frac{S_{450 \mu{\rm m}}}{S_{850 \mu {\rm m}}}\right> = 
\frac{e^{16.9/\langle{\rm T_{dust}}\rangle} - 
1}{e^{32.0/\langle{\rm T_{dust}}\rangle} - 
1}\left(\frac{850}{450}\right)^{3+\langle\beta\rangle}~~,
\end{equation}

\noindent where angle brackets denote spatial averages over each clump's 
area and the numerical exponents are the $h\nu/k$ terms from the Planck 
function.  Although we expect $\beta$ to vary somewhat with varying dust 
properties throughout massive star-forming regions, we presently have no 
means by which to quantify these variations.  For simplicity, we assume a 
spatially invariant value of $\langle\beta\rangle = 1.5$, consistent with 
the results of \citealt{dupac02}, who derived $\beta = 1.6\pm0.2$ for 
M17N, and \citealt{ss04}, who derived $\beta$ values ranging from
1.2 to 2 for massive clumps in NGC~7538.

	Using our assumed $\beta$ value and the flux ratios derived from 
the spectral index map, we can use Equation \ref{eq:tdust} to compute the 
mean dust temperature of each clump.  Again, due to the resolution 
mismatch between the \four\ and spectral index maps, this technique can 
only be applied to the \eight\ clumps.  Note that it is not possible to 
estimate temperatures for all of the \eight\ clumps.  The spectral index 
map contains only those pixels detected at $> 3 \sigma$ in both the \eight\ 
and \four\ images, so no spectral index can be derived for the thirteen 
\eight\ clumps which are not strongly detected at \four.  An additional ten 
clumps have measured flux ratios which are larger than the high 
temperature asymptotic value of the flux ratio from Eq. \ref{eq:tdust}.  
The presence of such high flux ratios probably reflects a breakdown in our 
assumption of a spatially invariant $\beta$, with unusually high flux 
ratios tracing regions of elevated $\beta$.  The estimated temperatures of 
the remaining 98 clumps are given in Table \ref{tab:m17850clumps}.  They 
range from 6 to 235~K, with a mean value of 33~K and a median value of 
20~K.  We derive the random uncertainties on these temperatures from the 
uncertainties in the gains, sky opacities, and the error beam fits.  The 
exponential form of Equation~\ref{eq:tdust} dictates that the width of the 
temperature uncertainty interval increases rapidly with temperature and 
that the upper uncertainty interval quickly exceeds the lower.  Typical 
random uncertainties on the estimated temperatures, which derive from the 
uncertainties in the gain and sky opacity calibrations, are $\pm 1.5$~K 
below 10~K, $\pm 5$~K between 10 and 20~K, $^{+10}_{-7}$~K between 20 and 
30~K, $^{+20}_{-10}$~K between 30 and 40~K, and $^{+50}_{-15}$~K above 
40~K.

	 Additional uncertainty enters via our choice of $\beta$: a clump 
with T$_{d}$ = 15~K for $\beta = 1.5$ would have T$_{d}$ = 11~K for $\beta 
= 2$ and T$_{d}$ = 21~K for $\beta = 1.2$.  The equivalent uncertainty 
range for a clump with T$_{d}$ = 30~K would be $\sim$16--90~K.  For these 
reasons, all of the temperatures above $\sim$30~K should be considered 
highly uncertain, indicating only that a core is probably ``hot''.  
Despite the uncertainties in their absolute values, the estimated 
temperatures still offer a means to rank the clumps by temperature.  One 
possible exception would be cases where two clumps having different 
temperatures lie within the same JCMT beam; we cannot rule out such 
occurrences.  To facilitate visualization of the distribution of clump 
temperatures, the clump symbols in Figure~\ref{fig:m17850map} are 
color-coded according to temperature.

\subsection{Clump Masses}

	Assuming that the dust emission is optically thin, the integrated 
dust continuum flux of a clump can be converted to a total clump mass via:

\begin{equation}
\label{eq:clumpmass}
M_{\rm clump} = \frac{S_{\lambda}^{\rm 
int}d^{2}}{\kappa_{\lambda}B_{\lambda}({\rm T}_{\rm dust})}~~,
\end{equation}

\noindent where $M_{\rm clump}$ is the total gas and dust mass of the 
clump, $S_{\lambda}^{\rm int}$ is the flux at wavelength $\lambda$ 
integrated over the clump boundary defined by \clfind, $d$ is the distance 
to the clump (taken to be 1.6~kpc), $\kappa_{\lambda}$ is the dust opacity 
per unit mass column density at wavelength $\lambda$, and 
$B_{\lambda}({\rm T}_{\rm dust})$ is the Planck function evaluated at 
temperature T$_{\rm dust}$.  Assuming, as previously, a spatially 
invariant $\beta = 1.5$, a gas-to-dust ratio of 100, and the \citet{h83} 
prescription for the dust opacity, $\kappa_{\lambda} = 0.1(250 \mu {\rm 
m}/\lambda)^{\beta}$, we obtain $\kappa_{850 \mu {\rm m}} = 
0.0087$~cm$^{2}$~g$^{-1}$ and $\kappa_{450 \mu {\rm m}} = 
0.031$~cm$^{2}$~g$^{-1}$.  This is the same prescription used in Paper~I 
to calculate the masses of the clumps in NGC~7538, and very closely 
mirrors those typically used in similar studies of the dusty clumps in 
low-mass star-forming regions (c.f. \citealt{dj2000}, \citealt{m01}).

	We adopt two separate approaches to the remaining parameter in 
Eq.~\ref{eq:clumpmass}, the dust temperature.  As a baseline approach, we 
make the simplifying assumption that all clumps are isothermal and 
characterized by a mean dust temperature of $\sim$30~K.  This is 
consistent with the dust temperature of $28\pm3$~K found by 
\citet{dupac02} using multi-wavelength PRONAOS dust continuum observations 
of M17N.  The masses calculated under this assumption range from 0.8 to 
120~$M_{\odot}$ for the \eight\ clumps, and from 1.8 to 160~$M_{\odot}$ for 
the \four\ clumps.  The median clump masses at \eight\ and \four\ are 12 
and 
13 $M_{\odot}$, respectively.  The total mass of the clumps detected at 
\eight\ and \four\ are 2900 and 2000~$M_{\odot}$, respectively.  In the 98 
cases where we were able to estimate the mean dust temperature of each 
\eight\ clump individually, we have used these temperatures to produce 
another estimate of their masses.  The clump masses so calculated range 
from 0.3 to 200~$M_{\odot}$, with a median value of 32~$M_{\odot}$.  
Clump masses calculated in both of these ways are given in 
Tables~\ref{tab:m17850clumps} and \ref{tab:m17450clumps}.  The masses 
calculated from the \eight\ fluxes are likely more accurate than those 
calculated from the \four\ fluxes, as the latter suffer from lower 
signal-to-noise and larger error beam corrections.  

	Like the temperatures, the masses are also affected by the choice 
of $\beta$.  A $\beta$ value of 2.0 would give \eight\ and \four\ masses 
which were 0.54 and 0.75 times the stated values, respectively.  For 
$\beta = 1.2$, the \eight\ and \four\ masses would be 1.4 and 1.2 times 
higher, respectively.  Note, however, that a change in the assumed 
constant value of $\beta$ does not change the shape of the clump mass 
function.

\subsection{Correlations With Signposts of Star Formation}
\label{sec:m17msx}

\begin{figure}
\begin{center}
\includegraphics[angle=270,width=6.0in]{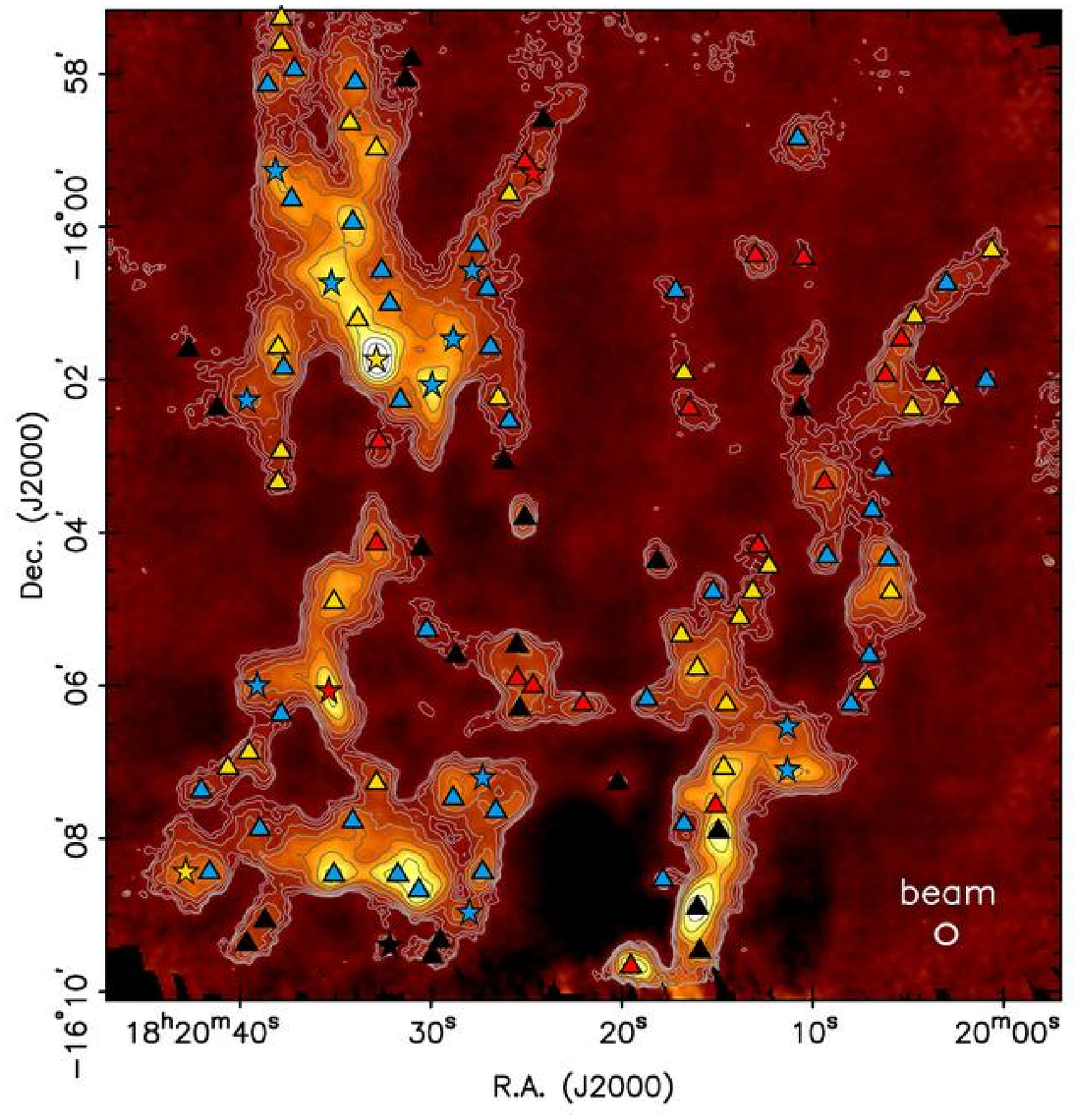}
\caption{Submillimeter continuum image of M17 at \eight, in halftone 
color with logarithmically spaced contours.  The contours begin at 
3~$\sigma$ (0.082 Jy beam$^{-1}$) and increase by
factors of 1.5.  Triangle and star symbols mark the clump peaks: stars 
indicate clumps which are spatially coincident with an MSX point source 
within their half-peak contour; triangles indicate clumps with no 
coincident MSX point source.  Symbol colors reflect a clump's estimated 
dust temperature: blue for T$_{\rm dust} \le 20$~K, yellow for 
20~K$ <$ T$_{\rm dust} \le$~40~K, red for T$_{\rm dust} >$~40~K, and 
black for those which lack a reliable temperature estimate.  The 
cross near M17N represents the position and positional uncertainty of the 
water maser detected by \citet{jgd81}.
\label{fig:m17850map}}
\end{center}
\end{figure}

\begin{figure}
\begin{center}
\includegraphics[angle=270,width=7.0in]{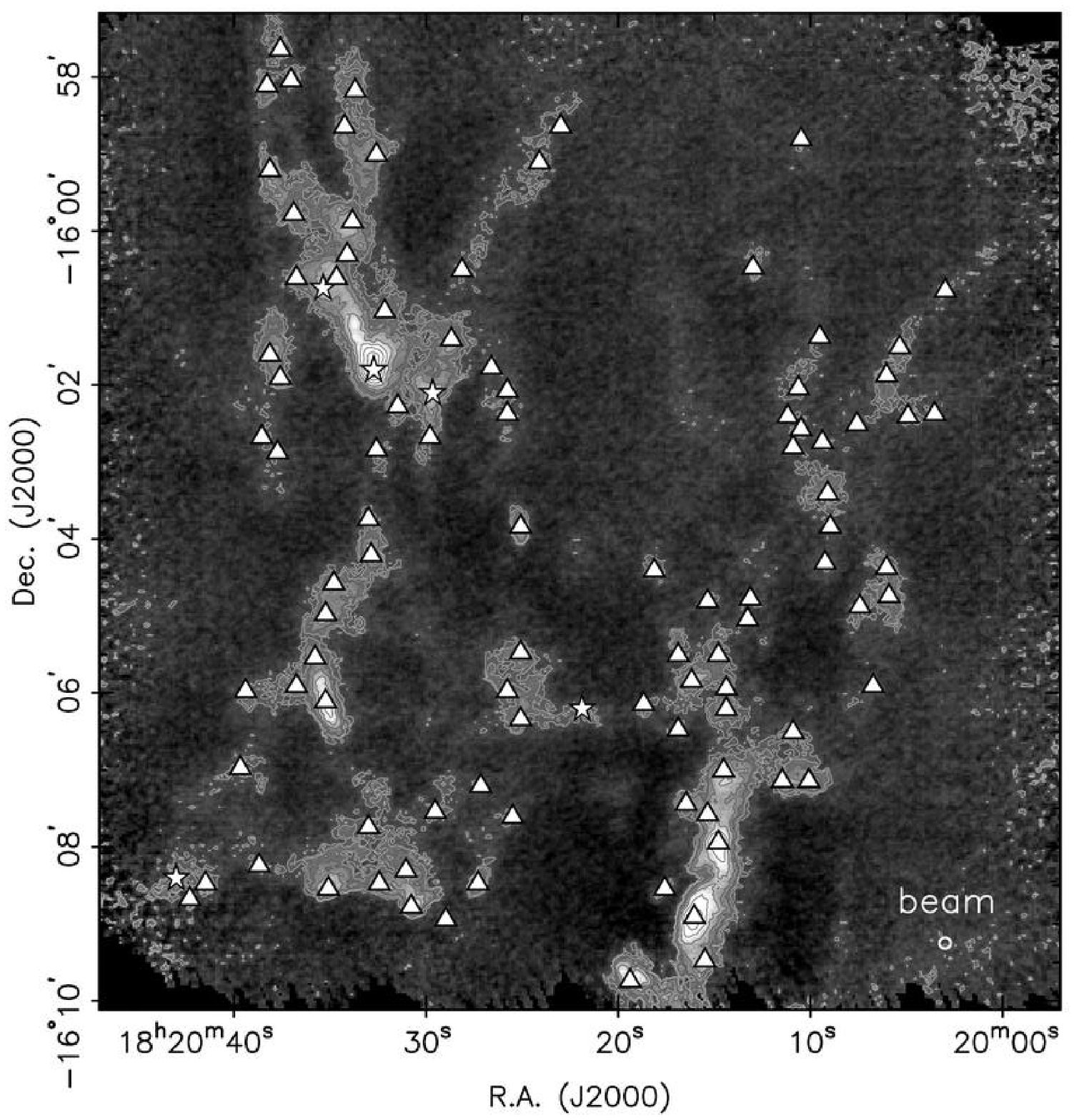}
\caption{Submillimeter continuum image of M17 at \four, in gray scale 
with logarithmically spaced contours.  The contours begin at 
3~$\sigma$ (0.96 Jy beam$^{-1}$) and increase by
factors of 1.5.  Triangle and star symbols mark the clump peaks: stars 
indicate clumps which are spatially coincident with an MSX point source 
within their half-peak contour; triangles indicate clumps with no 
coincident MSX point source.  \label{fig:m17450map}}
\end{center}
\end{figure}

	To assess the likely evolutionary states of the clumps in our map, 
we searched for spatial correlations between clumps and tracers of massive 
star formation, such as outflows, masers, and compact infrared sources.  
The mapped region of M17 does not appear to have been surveyed for outflow 
sources.  \citet{jgd81} surveyed our entire mapped region for 22~GHz water 
masers, locating only one, which is coincident with M17N within the 
positional uncertainties (see Fig.~\ref{fig:m17850map}).  To check for the 
presence of infrared sources, we compared our dust continuum clumps with 
the point source catalog from the Midcourse Space Experiment (MSX).  The 
number of MSX point sources lying within each clump's 50\% peak contour is 
given in Tables \ref{tab:m17850clumps} and \ref{tab:m17450clumps}.  The 
density of MSX point sources within the $> 3\sigma$ \eight\ emission region 
of our map is 50\% greater than that in the local field, so we estimate 
that approximately one third of the 37 sources within the $> 3\sigma$ 
boundary are likely to be physically associated with the M17 dust 
continuum clumps.  In total, we find 18 MSX point sources that are 
spatially coincident with 16 of our clumps, and we conclude that most of 
these MSX sources are probably physically associated with the 
submillimeter-emitting material.  Clumps which are coincident with MSX 
point sources are labeled with star symbols in Figures~\ref{fig:m17850map} 
and \ref{fig:m17450map}.

	We would expect clumps which have an embedded mid-infrared source 
to be among the hotter clumps, but this is only true in 2 of the 16 cases 
(clump SMM~51 in the center-north and SMM~98 in the southeast).  
Interestingly, 11 of the 16 clumps with coincident MSX point sources have 
estimated temperatures less than 20~K.  One possible interpretation is 
that these are cold clumps which have recently formed a hot, compact 
object.  Another possibility is that the flux ratio ceases to be a good 
measure of clump temperature in the presence of an embedded infrared 
source.  In some cases, the alignment of an MSX source with a clump 
probably coincidental.

	There are very few candidates in the literature for massive 
pre-stellar cores.  Of the 98 objects for which we were able to estimate 
temperatures, 39 have $T_{\rm dust} \leq 20$~K and are not coincident with 
an MSX point source.  The masses of these clumps range from 1.13 to 100 
$M_{\odot}$, indicating that they may become low, intermediate, or 
high-mass stars. While it is certainly possible that deeper infrared 
observations may reveal the presence of embedded sources within any of 
these clumps (see, for example, \citealt{young04}), these cold, apparently 
starless clumps represent the best candidates for massive pre-stellar 
cores in M17.

\section{THE MASS-RADIUS RELATIONSHIP OF THE M17 CLUMPS}
\label{sec:m17mvsr}

\begin{figure}
\begin{center}
\includegraphics[width=6.0in]{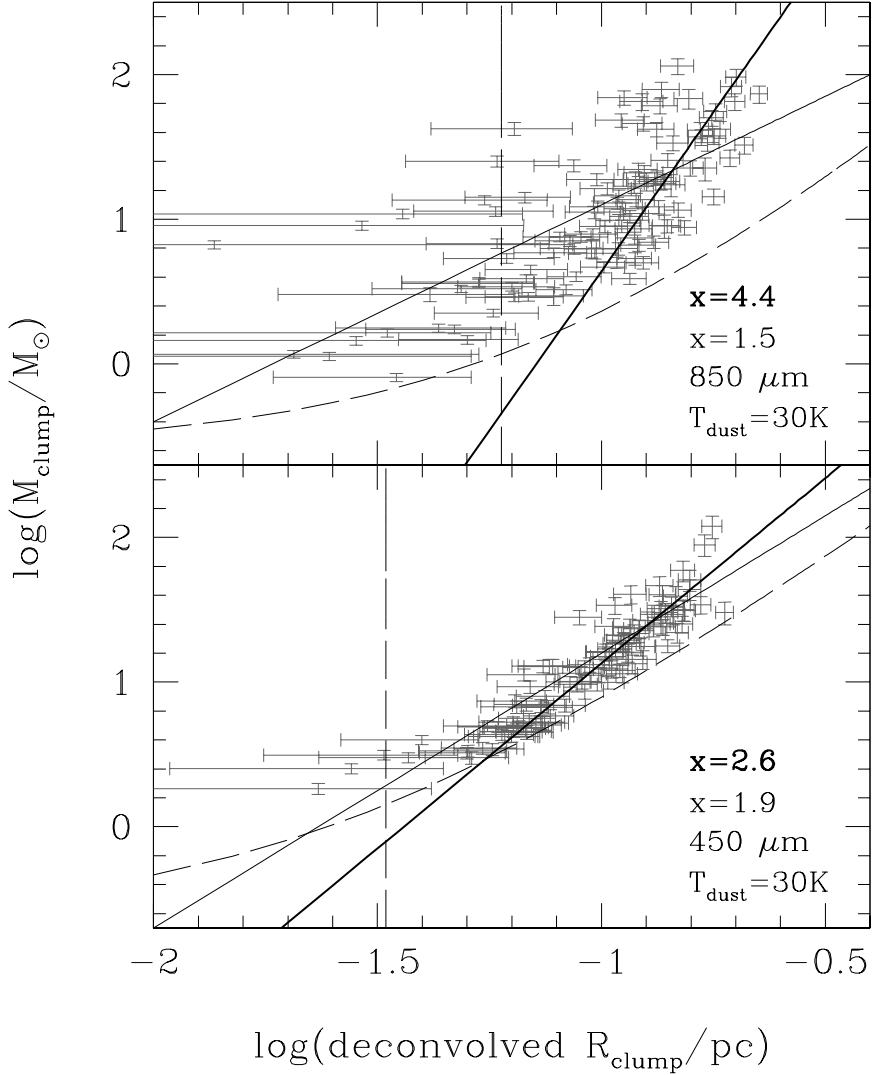}
\caption{The mass-radius relationship for three ensembles of M17 clumps,
with masses calculated assuming a uniform clump dust temperature of 30~K.  
The curving dashed lines represent the detection threshold.  The vertical
dashed lines represent the beam radius; clumps to the left of this line
are unresolved.  The plotted error bars represent the random mass
uncertainty and the uncertainty on the deconvolved radius assuming a
0.5\arcsec uncertainty in the beam diameter and a 2\arcsec (1 pixel)  
uncertainty in the clump radius before deconvolution.  The solid lines
represent fits to power laws of the form $M \propto R^{x}$; the thick line
shows a fit only to the resolved clumps and accounting for both their mass
and radius uncertainties, while the thin line fits all of the clumps and
ignores the uncertainties.  The values of the 
exponent, $x$, are given for each line in a 
correspondingly regular/bold typeface. \label{fig:m17_mvsr}} \end{center} 
\end{figure}

	By examining the relationship between the masses and radii of the 
M17 clumps, we may estimate the degree to which the sample is incomplete 
and determine something about the clumps' sources of support against 
collapse.  Because the majority of the clumps in our map are at least 
minimally resolved, the detection threshold must be expressed as a 
limiting surface brightness, not as a limiting integrated flux or an 
equivalent clump mass.  When the detection threshold is a limiting surface 
brightness, the clump sample may be incomplete to varying degrees at all 
masses: even a very massive clump may escape detection if its flux is 
spread sufficiently thinly.  The complex interaction between the limiting 
surface brightness and the actual, but unknown, distribution of clump 
masses makes it difficult to constrain the effects of incompleteness in 
submillimeter continuum clump studies.  The mass-radius plot helps to 
visualize the problem.  In Figure~\ref{fig:m17_mvsr}, we plot the mass of 
each clump versus its deconvolved effective radius.  For both the \eight\ 
(upper panel) and \four\ (lower panel) clumps, we use the masses 
calculated 
assuming a uniform clump temperature of 30~K and plot only the clumps with 
no coincident infrared source.

	We used these mass-radius plots to test whether the clump sample 
is essentially 100\% complete above the plotted detection threshold or 
whether the degree of completeness decreases as we approach the threshold 
from above.  The test consists of comparing the number of clumps extracted 
from a less sensitive version of our map to the number we would expect to 
extract based on the corresponding detection threshold drawn on 
Figure~\ref{fig:m17_mvsr}.  We made an \eight\ map using one fifth of the 
total integration time.  Drawing a corresponding detection threshold on 
the top panel of Figure~\ref{fig:m17_mvsr} (not shown), we find that we 
expect to lose 13 clumps in the less sensitive map.  After processing the 
less sensitive map in the same way as the original map, we find that we 
lose 16 clumps.  Thus, we conclude that the clump sample is close to 
complete above the detection threshold.  This suggests that, apart from 
potential sources of incompleteness such as chopping, the data are 
probably close to complete above the detection threshold.
	
	How should we interpret the distribution of clumps in the 
mass-radius plot?  If the clumps were an ensemble of critical Bonnor-Ebert 
spheres (i.e. thermally supported, pressure-bounded, self-gravitating, and 
on the verge of collapse), we would expect their mass-radius relationship 
to be well-fit by a power law of the form $M \propto R$ \citep{eb55,b56}.  
Studies of the mass-radius relationship of submillimeter clumps in 
low-mass star-forming regions such as $\rho$~Oph and Orion B have found 
good agreement with the $M \propto R$ power law \citep{m01}.  Similarly, 
if the clumps were an ensemble of critical, non-thermally supported 
spheres, such the logatropic spheres of \citet{mp96}, then they ought to 
obey a relationship more like $M \propto R^{2}$.  In Paper~I, we found 
that the \eight\ clumps in NGC~7538 were best fit by a power law, $M 
\propto R^{x}$, with exponent $x = 2.1 \pm 0.1$, suggesting that they are 
primarily turbulently supported.  Both of these results are somewhat 
surprising because, taken at face value, they imply an implausible 
scenario in which most of the clumps hover somewhere near criticality.

	In the present work, refinements to our fitting technique require 
a more complex interpretation of the M17 clump mass-radius relationship.  
If we fit the mass-radius relationship of all of the clumps, taking no 
account of their radius uncertainties and making no distinction between 
resolved and unresolved clumps, we find results similar to those found in 
NGC~7538 (Paper~I): the \eight\ and \four\ mass-radius relationships have 
exponents of $x = 1.5 \pm 0.1$ and $x = 1.9 \pm 0.1$, respectively.  
However, if we exclude all those clumps with deconvolved radii smaller 
than the beam radius (those to the left of the vertical dashed line in 
Fig.~\ref{fig:m17_mvsr}) and use both the mass and radius uncertainties in 
computing the fit, we find $x = 4.4 \pm 0.2$ for the \eight\ clumps and $x 
= 2.6 \pm 0.1$ for the \four\ clumps.  The steepness of the \eight\ 
mass-radius relationship in M17 is surprising, though it depends 
sensitively on the precision with which we can measure the radii of the 
smallest clumps.  In principle, the \four\ data, with twice the linear 
resolution of the \eight\ data, should produce a more reliable value of the 
exponent, $x$.  The $x$ value of $2.6 \pm 0.1$ is much closer to those 
obtained in similar studies of low-mass star-forming regions.  It is 
unclear, however, whether those studies also accounted for the potentially 
significant radius uncertainties and therefore whether their results are 
comparable with ours.  Further uncertainties may arise due to 
incompleteness and unresolved substructure in some clumps.  Neither effect 
can be reliably accounted for at present.  Assuming that the value $x = 
2.6 \pm 0.1$ from the \four\ mass-radius relationship is correct, it 
indicates that the clumps are intermediate between constant surface 
density ($x=2$) and constant volume density ($x=3$), and that they are 
non-thermally supported.

\section{THE M17 CLUMP MASS FUNCTION} 
\label{sec:m17massfunc}

\subsection{Fitting the M17 Clump Mass Function}
\label{sec:m17mffit}

	Astrophysical mass functions are commonly constructed in two ways:
differential ($\Delta N/\Delta M$), in which the data are binned by mass,
and cumulative ($N(>M)$), in which the data are not binned.  We consider
both forms, as each has unique merits.  The binning process inherent in
the differential mass function (DMF) naturally averages out errors and
enables straightforward accounting of uncertainties, but the relatively
arbitrary assignment of bin widths and centers introduces bias and can
obscure physically significant features of the mass function.  The
cumulative mass function (CMF) does not average out errors, but it also
does not lose information to binning, an effect which has been found
problematic in studies of the stellar mass function
\citep{scalo98,apellaniz}.  Lack of binning makes the treatment of
uncertainties somewhat more complicated in the case of the CMF.  In the
limit of very many objects ($N \rightarrow \infty$), the differential mass
function might be preferable.  However, studies of the clump mass function
are typically in the relatively small $N$ regime, where neither form of
the mass function should be considered definitive and careful
consideration of both forms is warranted.

\begin{figure} 
\begin{center} 
\includegraphics[width=5in]{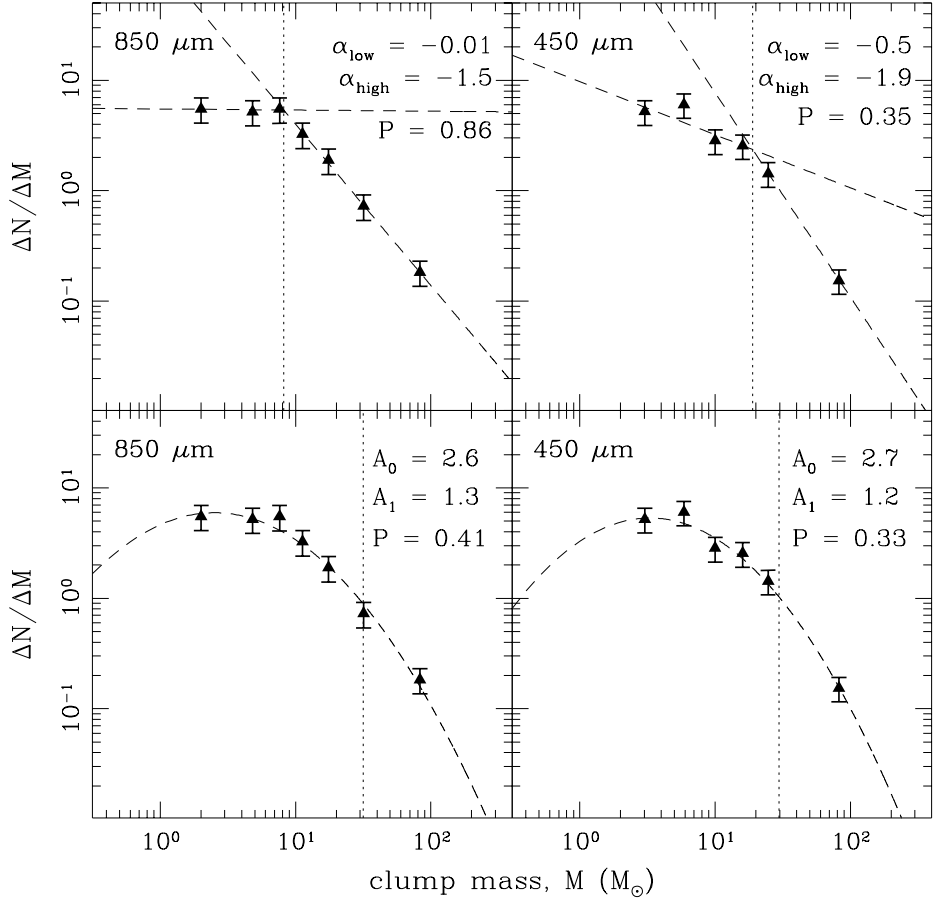}

\caption{Differential mass function for the M17 starless clumps detected
at \eight\ (left) and \four\ (right), with clump masses calculated assuming
a constant dust temperature of 30~K.  In the top row, the mass functions
are fitted with a broken power law (\emph{dashed lines}), whose break
(\emph{dotted lines}) is a parameter of the fit.  The best fit exponents
of the two power laws are shown in the upper right corner of each panel.  
In the bottom row, the data are fitted with a lognormal shape
(\emph{dashed curve}), defined by Eq.~\ref{eq:logdmf}, with best fit
parameters as shown.  The vertical dotted line in the lower panels
represents the mean mass derived from the lognormal fit.  Values for all
the fitted parameters, with uncertainties, are given in
Table~\ref{tab:m17params}.\label{fig:m17dmfs}}
\end{center} 
\end{figure}

\begin{figure} 
\begin{center} 
\includegraphics[width=5in]{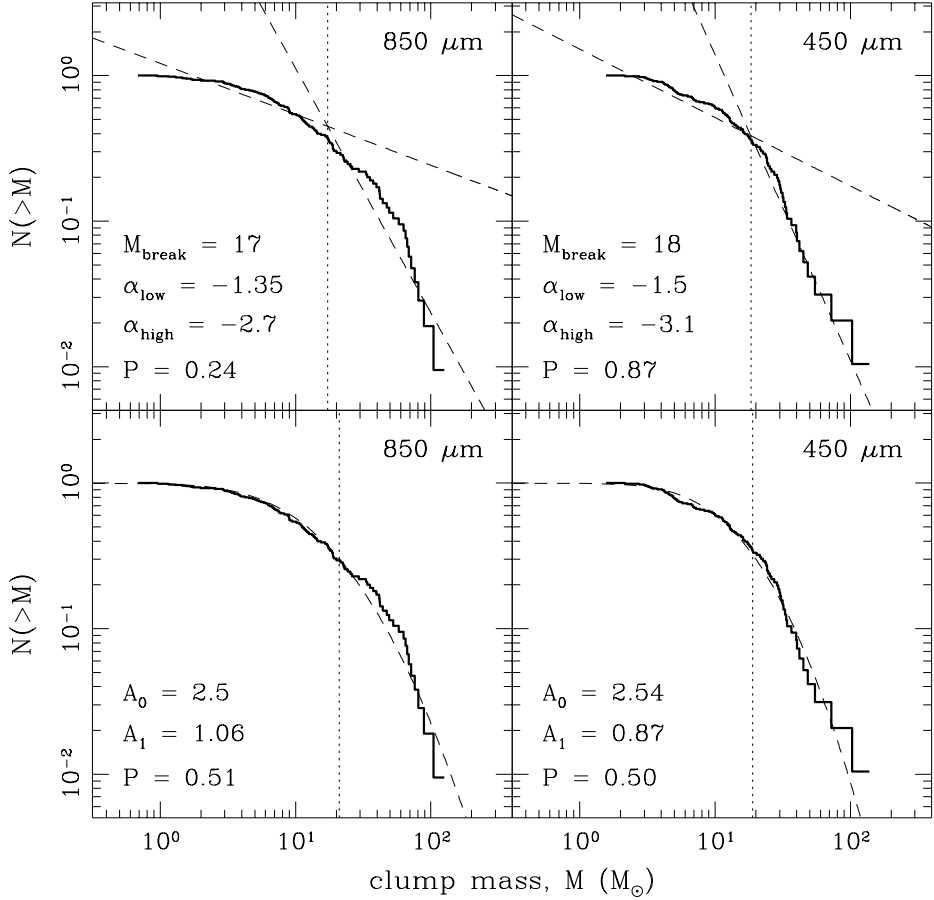}

\caption{Cumulative mass function for the M17 starless clumps detected at
850 (\emph{left}) and \four\ (\emph{right}), with clump masses calculated
assuming a constant dust temperature of 30~K.  In the top row, the mass
functions are fitted with broken power laws (\emph{dashed lines}), whose
break points are also parameters of the fit (\emph{dotted lines}).  In the
\eight\ CMF, three power laws (not shown) are required to obtain a good
fit.  For consistency, we show only the fit with two power laws.  The
exponents of the best fit power laws are shown in the lower left corner of
each panel.  In the bottom row, the data are fitted to the CMF
corresponding to a lognormal DMF (see Eq. \ref{eq:logcmf}), with best fit
parameters as shown.  The vertical dotted line in the lower panels
represents the mean mass derived from the lognormal fit.  Values for all
the fitted parameters, with uncertainties, are given in
Table~\ref{tab:m17params}.  The $P$ values are goodness-of-fit measures,
with values $\gtrsim0.1$ indicating a good fit (see \S\ref{sec:m17mffit}).  
\label{fig:m17cmfs}} \end{center} \end{figure}

	We have constructed both forms of the mass function for the M17 
clumps detected in each waveband.  Figure~\ref{fig:m17dmfs} shows the 
differential mass function of the M17 clumps at 850 and \four.  To 
minimize uncertainties caused by the choice of bin width, we have followed 
the prescription of \citet{apellaniz} and used a variable bin width with a 
constant number of clumps per bin.  Figure~\ref{fig:m17cmfs} shows the 
cumulative mass function of the M17 clumps in both wavebands.  Because we 
are primarily interested in the mass function of pre-stellar clumps, we 
exclude from the analysis clumps which are spatially coincident with MSX 
point sources (as per the discussion of \S\ref{sec:m17msx}).  In the 
literature on low-mass pre-stellar cores, the differential version of the 
core mass function is typically fit using a double power law of the form:

\begin{equation}
\frac{\Delta N}{\Delta M} \propto \left\{ \begin{array}{rl}
 M^{\alpha_{\rm low}} & ,~M < M_{\rm break} \\
 M^{\alpha_{\rm high}} & ,~M \geq M_{\rm break} 
 \end{array} \right.
\label{eq:dmfeq}
\end{equation}

\noindent where $\alpha_{\rm low}$, $\alpha_{\rm high}$, $M_{\rm break}$,
and a normalization factor may all be fitted parameters.  For the
cumulative form of the mass function, the double power law fit takes the 
form:

\begin{equation}
N(>M) \propto \left\{ \begin{array}{rl}
 M^{\alpha_{\rm low}+1} & ,~M < M_{\rm break} \\
 M^{\alpha_{\rm high}+1} & ,~M \geq M_{\rm break} 
 \end{array} \right.
\label{eq:cmfeq}
\end{equation}

\noindent where the symbols have the same meanings as in
Equation~\ref{eq:dmfeq}.  In both forms, the exponent of the Salpeter mass
function would be $\alpha_{\rm high} = -2.35$ \citep{sal55}.  In the top
rows of Figures~\ref{fig:m17dmfs} and \ref{fig:m17cmfs}, we show fits to
double power laws of the forms given in Eqs.~\ref{eq:dmfeq} and
\ref{eq:cmfeq}.  In each fit, $\alpha_{\rm low}$, $\alpha_{\rm high}$,
$M_{\rm break}$, and the normalization factor are unconstrained
parameters, fitted simultaneously.

	To fit the DMF, we used a standard nonlinear Levenberg-Marquardt 
least squares algorithm \citep{press}, with Poisson error bars, as shown, 
used for the weights.  Poisson error bars are not suitable for fitting the 
CMF, however, because it is not binned.  Instead, we used a Monte Carlo 
technique to compute the best fit parameters, their uncertainties, and a 
measure of the goodness-of-fit.  For each data set, we generate $10^{5}$ 
synthetic data sets by randomly varying the mass of each clump within a 
Gaussian probability envelope whose mean and standard deviation are equal 
to the observed clump mass and its random uncertainty, respectively.  
Each of the $10^{5}$ synthetic data sets is then fit using the 
Levenberg-Marquardt nonlinear least-squares algorithm.  To compute 
approximate weights used in the calculation and minimization of 
$\chi^{2}$, we begin by noting that, to a good approximation, the observed 
random mass errors scale as $\sigma_{M} \propto M$.  If we approximate the 
CMF by a power law $N(>M) \simeq AM^{-x}$, then we find the following by 
partial differentiation:

\begin{equation}
 \sigma_{N} \simeq A(-x)M^{-x-1}\sigma_{M} \propto N~~.
\end{equation}

\begin{figure}
\begin{center}
\includegraphics[width=\columnwidth]{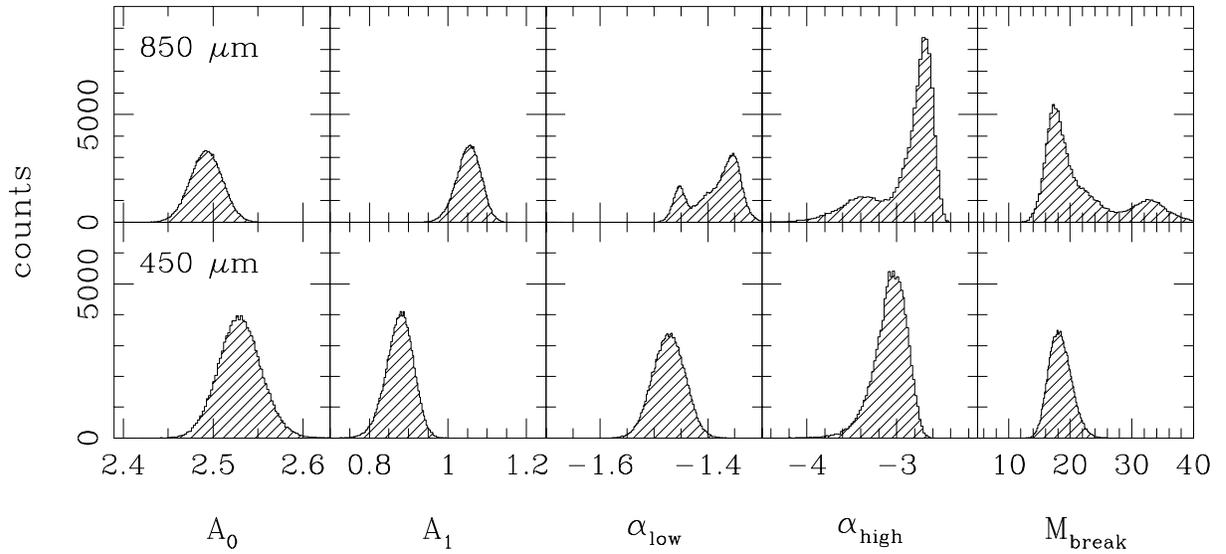}

\caption{Histograms of the best fit parameters derived from lognormal and
double power law fits to 10$^{5}$ realizations of the M17 cumulative clump 
mass
function within the random uncertainties on each clump mass.  Best fit
parameter distributions are shown for both the \eight\ CMF (\emph{top row}) 
and the \four\ CMF (\emph{bottom row}).  In general, the distributions are 
close to normal, except for the double power law fits to the \eight\ CMF, 
which are bimodal.\label{fig:m17_bfph}}
\end{center}
\end{figure}

\noindent Thus, we take the statistical weight of each point in the mass
function to be $1/\sigma_{N}^{2} \simeq 1/N^{2}$, which accords with our
intuition that clumps of higher masses (lower $N$ in the CMF) should be
more heavily weighted.  Using this technique, we fit each of the $10^{5}$
synthetic data sets, plot histograms of the best fit parameters, and
report the peak of each parameter distribution as the best fit to the
observed mass function.  The uncertainties on the CMF fits are taken to be
the 95\% (2 $\sigma$) confidence intervals on either side of the best fit 
parameters.  Histograms of the fitted parameters from the Monte Carlo
distributions are shown in Figure~\ref{fig:m17_bfph}; where the distribution
of parameters is nearly symmetric, we quote equal upper and lower 
uncertainties.  The best fit parameters calculated according to
this Monte Carlo technique are summarized in Table~\ref{tab:m17params}.

	Both forms of the \four\ mass function are well fit over most of 
their mass range by two power laws with a break at $\sim 19 M_{\odot}$.  
The \eight\ DMF is also well fit by two power laws with a slightly lower 
break mass ($9 \pm 3~M_{\odot}$).  The \eight\ CMF, however, is not very 
well fit by a double power law, as evidenced by the bimodal parameter 
distribution which results when we attempt a double power law fit (see 
Fig.~\ref{fig:m17_bfph}).  The higher of the two peaks in each parameter 
distribution represent the best fit to a double power law with a break 
mass at roughly the same position as the break found in the \four\ mass 
function ($\sim 19 M_{\odot}$).  The lower peak demonstrates that a third 
power law with a steeper slope at masses above $\sim 40 M_{\odot}$ is 
required to produce a good fit.  A mass function might be well fit by two 
power laws because it is incomplete at low masses or because there is a 
real physical break in the distribution of clump masses.  However, that a 
a power law fit in three segments is required to produce a good fit 
suggests that the power laws are merely approximating an unknown 
continuous distribution.  We consider this possibility next.

	As will be discussed more extensively in Paper~III, the lognormal
distribution is a likely candidate for a continuous clump mass
distribution.  The Galactic field star IMF can be fit by a lognormal mass
distribution (\citealt{ms79,chabrier03}, but see \citealt{scalo98} for a
dissenting view), as can the mass function of pre-stellar clumps obtained
from hydrodynamic simulations \citep{klessen01}.  Moreover, several
analytic studies of the functional form of the clump mass function have
found the lognormal distribution to be a viable candidate
\citep{larson73,z84,af96}.  In the bottom row of Figure~\ref{fig:m17dmfs},
we fit the M17 DMFs with a lognormal mass function of the form:

\begin{equation}
\frac{\Delta N}{\Delta M} = \frac{1}{A_{1}
\sqrt{2\pi}M}~\mbox{exp}\left[-\frac{(\mbox{ln}M - 
A_{0})^{2}}{2A_{1}^{2}}\right]~~.
\label{eq:logdmf}
\end{equation}  

\noindent Similarly, in the bottom row of Figure~\ref{fig:m17cmfs}, we fit 
the CMFs with the cumulative form of the above clump mass distribution,

\begin{equation}
N(>M) = \frac{1}{2}\left[1 - \mbox{erf}\left(\frac{\mbox{ln}M - 
A_{0}}{\sqrt{2}A_{1}}\right)\right]~~.
\label{eq:logcmf}
\end{equation}  

\noindent The best fit values of $A_{0}$ and $A_{1}$ accompany each plot.  

	Interestingly, both the 850 and \four\ CMFs are well fit by a 
lognormal clump mass distribution.  The cumulative distribution function 
of $\chi^{2}$ values from our Monte Carlo simulations of the data allow us 
to determine the probability, $P$, of obtaining by chance fits as poor as 
our best fits.  The $P$ value for each fit is shown in 
Figure~\ref{fig:m17cmfs}, where $0<P<1$.  In all cases, both power law and 
lognormal mass distributions generate acceptable fits ($P \gtrsim 0.1$; 
\citealt{press}).  As suggested above, the continuous lognormal 
distribution provides a better fit to the \eight\ CMF than does the double 
power law.  Conversely, the double power law provides a better fit to the 
\four\ CMF than the lognormal distribution, though both fits are good.  
Based on these results, we suggest that both the lognormal distribution 
and the double power law are viable functional forms for the mass function 
of the pre-stellar clumps in M17.  It is interesting to note that the 
lognormal distribution produces a good fit to the CMFs using only 2 
parameters (mean mass and width), whereas the double power law requires 4 
parameters (two exponents, a break point, and a normalization) to produce 
fits of comparable quality.

	Does the quality of the lognormal fits reflect the true 
distribution of the clump masses, or is it merely coincidence that an 
intrinsically power-law mass function is reasonably well fit by a 
lognormal distribution?  Experimentation with synthetic data sets quickly 
reveals that, given sufficient freedom in selecting values of $\alpha_{\rm 
low}$, $\alpha_{\rm high}$, and $M_{\rm break}$, it is possible to 
generate an intrinsically double power law mass distribution whose CMF is 
significantly better fit by a lognormal distribution than by a double 
power law.  However, the range of parameters for which this is the case 
are not necessarily reflective of those which characterize real 
star-forming regions.  We would like to determine, in the specific case of 
M17, the probability that a double power law mass distribution with 
plausible parameters might appear better fit by a lognormal distribution. 
To assess this probability, we generated $10^{5}$ synthetic 850 and \four\ 
clump mass functions with exactly the mass distributions of our best fit 
double power laws.  The CMF of each synthetic data set was fit with both a 
double power law and a lognormal distribution.  We find that the residuals 
for the lognormal fit exceed those of the double power law fit in 90\% of 
the trials with the synthetic \eight\ CMFs and 82\% of the trials with the 
\four\ CMFs.  Thus, it appears unlikely that the good quality of our 
lognormal fits reflects a misrepresentation of an intrinsically double 
power law mass function in M17.  Paper~III will include a more 
comprehensive discussion of the uncertainties inherent in determining the 
functional form of the clump mass function.

	Another interpretation of the lognormal fits to the M17 clump mass 
function is that severe incompleteness at their low-mass ends distorts 
what would otherwise be a single power law distribution.  Incompleteness 
would produce a mass function with a shallow low-mass end, potentially 
causing the lognormal distribution to produce the best fit.  However, 
given that the instruments, data reduction techniques, and analytical 
methods are very similar in our studies of high-mass star-forming regions 
and previous studies of low-mass regions, we believe that incompleteness 
should affect all of the measured mass functions similarly.  Comparison of 
the mass-radius plots in our Figure~\ref{fig:m17_mvsr} with Figure 4 of 
\citet{m01} shows that, in both cases, the distribution of clumps adheres 
closely to the detection threshold, suggesting that both studies should be 
similarly affected by incompleteness.  If the clump mass function in 
low-mass star-forming regions were severely incomplete at the low-mass 
end, it would spoil good the agreement between the clump and stellar mass 
functions (see \S\ref{sec:compstellar}).  The corrected clump mass 
function would then show a considerable overabundance of pre-stellar 
clumps below $\sim 1~M_{\odot}$ compared to the number of low-mass stars 
in the IMF.  Thus, we have no reason to believe that incompleteness should 
dramatically alter the shape of the clump mass function in low-mass 
star-forming regions, nor that incompleteness has widely disparate effects 
on studies of low- and high-mass star-forming regions.  Hence, we cannot 
conclude that the quality of the lognormal fits is due to incompleteness.  
The mass function of M17 may indeed have a lognormal shape with a peak 
above $1~M_{\odot}$.

\subsection{Comparison to the CO Clump Mass Function}

	Studies of CO clumps in star-forming regions have typically found 
mass functions which obey shallower power laws than those of submillimeter 
continuum clumps spanning the same mass range \citep{kramer98}.  In 
Paper~I, we compared the power law exponents of both CO and submillimeter 
continuum mass functions and found them to be genuinely discrepant.  The 
evidence from M17 further supports this conclusion.  \citet{kramer98} 
found that the mean power-law exponent of the differential mass functions 
of CO clumps in 7 star-forming regions whose clumps span 
$10^{-4}$--$10^{4} M_{\odot}$ is $-1.69\pm0.02$.  \citet{sg90} found that 
the DMF of CO clumps in the M17SW region, which adjoins but does not 
overlap the region we mapped, is well fit over the mass range \mbox{M 
$\simeq10$--2000~$M_{\odot}$} by a single power law with exponent 
$-1.72\pm0.15$.  In M17, where the mass function of submillimeter 
continuum clumps spans a compatible range of $\sim 10$--$120 M_{\sun}$, 
we find a mean power law exponent of $\alpha_{\rm high} = -2.3\pm0.4$.  
Again, the submillimeter continuum and CO mass functions appear to 
disagree.  In Paper~III, we will present an analysis of the functional 
form of the submillimeter continuum mass function which suggests that it 
is not well fit by a single power law, as the CO mass functions clearly 
are.  The reason for the discrepancy between the CO and submillimeter 
continuum mass functions is not yet clear.  To help resolve this nagging 
issue, it would be illuminating to compare the mass functions of objects 
extracted from CO and submillimeter continuum maps of identical regions, 
at comparable spatial resolutions, using the same clump extraction 
technique.  For the moment, the issue remains unresolved.

\subsection{Comparison to the Stellar IMF and the Mass Function of Low-Mass Clumps}
\label{sec:compstellar}

	The masses of the M17 clumps span the high-mass end of the stellar
mass range.  The masses computed assuming a constant dust temperature
range from 0.8 to 160~$M_{\odot}$, although only a few clumps have masses
$> 100$~$M_{\odot}$.  Their deconvolved radii span the range
\mbox{$\sim$0.02--0.2}~pc, or \mbox{$\sim$4000--40000~AU}.  Although the
large distance to M17 makes it difficult to match the linear resolution
and mass sensitivity of similar studies of low-mass star-forming regions,
there is significant overlap.  For example, in their study of the dust
continuum clumps in the Orion B complex, \citet{dj2001} found clumps with
masses and radii in the ranges \mbox{0.06--30}~$M_{\odot}$ and
\mbox{3100--20000~AU}, respectively.  The properties of our clumps also
overlap, to differing extents, with those found in the submillimeter
continuum studies of $\rho$~Oph \citep{man98,dj2000}, the Lagoon Nebula
\citep{tot}, and NGC~7538 (Paper~I), among others.  Based on their masses,
sizes, and estimated temperatures, it is reasonable to suggest that some
fraction of the ``clumps'' in M17 are in fact star-forming ``cores'' which
may be the direct precursors of individual stars or small multiple
systems.  Thus, as has been done in studies of the clump mass function in
low-mass star-forming regions, it is reasonable to compare the M17 clump
mass function to the mass function of stars in young clusters.  Ideally,
because we do not know how many stars may form in any given clump, we
would compare our clump mass function to the IMF of stellar \emph{systems}
in young clusters.  The IMF of stellar systems differs from the IMF of
individual stars by corrections for unresolved binaries and higher order
multiples.  In practice, because these corrections are difficult to
determine accurately, many published ``stellar'' IMFs include unresolved
systems \citep{kroupa}.

	The exact form of the stellar mass function in young clusters is 
not yet well established, particularly for cluster ages $\lesssim 
10^{6}$~yr \citep{chabrier03}.  The present-day mass function of young 
cluster stars can be represented by a lognormal distribution below $\sim 
1$~$M_{\odot}$ and a power law above, though whether there may be multiple 
power law segments at higher masses is not clear.  (The IMF of Galactic 
field stars, for example, is well fit by three power law segments above 
$1~M_{\odot}$; \citealt{chabrier03}.)  Where a single high-mass power law 
is assumed, its exponent is found to be roughly equal to the Salpeter 
value, i.e. $\Delta N/\Delta M \propto M^{-2.3}$ 
\citep{sal55,kroupa,chabrier03}.  Measurements of $\alpha_{\rm high}$ for 
the stellar mass function of young clusters exhibit substantial variation, 
ranging at least between -1.5 and -3.  \citep{kroupa} have suggested that, 
because most measured $\alpha_{\rm high}$ values are approximately 
normally distributed around -2.3, the spread may be due to ordinary 
measurement variation.  However, other authors favor a steeper value for 
$\alpha_{\rm high}$, perhaps closer to -3, for massive stars 
\citep{sr91,casassus}.  Studies of the pre-stellar core mass function in 
low-mass star-forming regions have shown peak masses ($\lesssim 
0.5~M_{\odot}$) and power law exponents for $M_{\rm clump} \gtrsim 
1$~$M_{\odot}$ which agree well with those found in the young cluster 
stellar IMF \citep{ts98,man98,dj2000,dj2001,m01}.  This agreement between 
the core mass function and the stellar IMF has been widely interpreted as 
suggesting that the cores are the immediate precursors of stars, and that 
the core mass function will translate smoothly into a similarly-shaped 
stellar IMF.

	Our four measures of $\alpha_{\rm high}$ for the M17 clumps have 
much the same characteristics as measures of $\alpha_{\rm high}$ for the 
stellar IMF.  As shown in Table~\ref{tab:m17params}, the $\alpha_{\rm 
high}$ values range between -1.5 and -3.0, with a mean of -2.3$\pm$0.4, 
consistent with the Salpeter value.  As with measures of $\alpha_{\rm 
high}$ for the stellar IMF, variations among the measures of $\alpha_{\rm 
high}$ for the M17 clump mass function are likely attributable to ordinary 
measurement variation.  In particular, the uncertain placement of the 
break between the two power laws can strongly influence the derived power 
law exponents.  Even though we find a mean power-law exponent which agrees 
with the Salpeter value, this does not necessarily mean that the M17 clump 
mass function is well-represented by a single power law at high masses.  
Indeed, as discussed in \S\ref{sec:m17mffit}, the high-mass end of the 
\eight\ CMF is best fit by 2 separate power law segments.  Thus, the 
apparent agreement with the Salpeter power law may be misleading: in 
particular, our good lognormal fits suggest that the structure of the 
clump mass function is probably somewhat more complicated than this 
agreement would suggest.  

	The mass function of high-mass stars in young clusters is 
typically assumed to have a power law form, so lognormal fits to it are 
scarce.  The IMF of Galactic field stars is presumed lognormal below $\sim 
1~M_{\odot}$, with a peak near 0.1~$M_{\odot}$, but it may also be well 
fit by a lognormal distribution at masses above 1~$M_{\odot}$ 
(\citealt{ms79,chabrier03}, although see \citealt{scalo98}).  As mentioned 
above, studies of the clump mass function in low-mass star-forming regions 
have typically found that it peaks well below $1~M_{\odot}$, at several 
0.1~$M_{\odot}$ \citep{ts98,man98,dj2000,dj2001,m01}.  The range of clump 
masses probed by these studies typically ranges from a few 
0.01~$M_{\odot}$ to as much as 30~$M_{\odot}$.  All of our 
lognormal fits to the M17 clump mass function (both wavelengths, both 
forms of the mass function) indicate that it peaks above 1~$M_{\odot}$.  
The fitted peak masses, shown in Table~\ref{tab:m17params}, range from 2 
to 5.7~$M_{\odot}$ with a mean of $4.2\pm0.7~M_{\odot}$, while the clump 
masses themselves span the range $\sim$1--150~$M_{\odot}$.  This suggests 
that the M17 mass function is not simply the extension to higher masses of 
the clump mass function seen in low-mass star-forming regions.  This 
hypothesis is supported by our finding that the clump mass functions in 
both M17 and NGC~7538 (Paper~I) are clearly best fit by more than one 
power law above 1~$M_{\odot}$.  The position of the peak mass in each 
clump mass function is affected by incompleteness, though the extent of 
this effect is difficult to assess.  Furthermore, the differing 
assumptions made in each study about the clump temperatures and dust 
emissivities affect the derived value of the peak clump mass.  Subject to 
these caveats, we suggest that the clump mass function in high-mass 
star-forming regions may be a \emph{scaled-up} version of the low-mass 
clump mass function.  The two may have a common form (perhaps lognormal) 
but different intrinsic scales, as characterized by their differing peak 
masses.
 	
\section{SUMMARY}
\label{sec:m17summary}

	We have produced \eight\ and \four\ SCUBA continuum maps of an
approximately 12\arcmin$\times$12\arcmin~region of the M17 star-forming
region, including the source M17N.  We used \clfind\ to extract the clumps
from each map, making appropriate corrections for radio continuum
contamination and the JCMT error beam.  We computed the mean spectral
index, $\alpha$, for each clump and thereby estimated their mean dust
temperatures.  We computed the mass of each clump using both the
calculated temperatures and a universal mean temperature of 30~K.  We used
MSX data to make a preliminary determination as to which clumps have
embedded sources and which are starless.  Our main findings can be
summarized as follows:

\begin{enumerate}

\item We identify 121 clumps at \eight\ and 101 at \four.  Of the \eight\
clumps, 105 are found not to be associated with an MSX point source within
their half-peak flux contours.  Of the \four\ clumps, 96 are not coincident 
with an MSX point source.  Assuming a constant dust temperature of 30~K, the
\eight\ clump masses span the range 0.8--120~$M_{\odot}$ and the \four
clump masses span the range 1.8--160~$M_{\odot}$.

\item By combining the \four\ and \eight\ data for each clump with the 
assumption of a constant dust emissivity index, $\beta$, we derive a mean 
spectral index, $\alpha$, and dust temperature, $T_{\rm dust}$ for each 
clump detected at \eight.  The spectral indices range from 0.45 to 3.45, 
with a mean of 2.74.  The estimated dust temperatures range from 6 to 
235~K, with a mean of 33~K and a median of 20~K.  As in Paper~I, we find a 
correlation between high spectral index and high submillimeter continuum 
flux.

\item Fitting the mass-radius relationship of the clumps to a power law, 
$M \propto R^{x}$, we find $x = 4.4\pm0.2$ for the \eight\ clumps and 
$x=2.6\pm0.1$ for the \four\ clumps.  In both wavebands, the distribution 
of clumps in mass-radius space adheres fairly close to the detection 
threshold, which is of the form $M \propto R^{2}$, making it difficult to 
assess the effects of incompleteness.  If the observed mass-radius 
relationship accurately reflects the characteristics of the M17 clumps, we 
suggest that it implies that they derive significant support from 
nonthermal motions.

\item We have identified 39 ``cold'' clumps ($T_{\rm dust} \lesssim 20$~K)  
which are not spatially coincident with an MSX point source.  We suggest
that these are the best candidates for very young massive pre-protostellar
cores in the mapped portion of M17.

\item We have produced both differential ($\Delta N/\Delta M$) and
cumulative ($N(>M)$) mass functions for the ``starless'' clumps detected
in each waveband.  We fit each mass function with both a segmented power
law and a lognormal distribution.  All forms of the mass function are well
fit by a double power law except the \eight\ CMF, which would require three
power law segments for a good fit.  The mean power-law exponent of the
high-mass end of the clump mass function is $\alpha_{\rm high} =
-2.4\pm0.3$.  On the same scale, the Salpeter value would be $\alpha_{\rm
high} = -2.35$.

\item We suggest the lognormal distribution as a candidate continuous form 
for the clump mass function in M17.  The lognormal distribution fits the 
M17 mass function as well as or better than the commonly-used double power 
law, and with fewer fitted parameters (2 instead of 4).  The lognormal 
distribution also eliminates the need to invoke a physically unmotivated 
third power law segment to obtain a good fit to the \eight\ CMF.  A 
lognormal mass function would be consistent with the view of high-mass 
star formation as an inherently clustered, and therefore highly stochastic 
phenomenon (e.g. \citealt{af96}).  The mean peak mass of the fitted 
lognormal distributions is $3.8\pm0.7~M_{\odot}$, which assumes a dust 
emissivity of $\beta$ = 1.5 and a mean dust temperature of $T=30$~K.  This 
mass is 
substantially higher than the peak mass obtained from fits to the stellar 
IMF and low-mass clump mass function, which is typically a few tenths of a 
solar mass.

\item We suggest that the clump mass function in massive star-forming
regions may be a scaled-up version of the mass function seen in low-mass
star-forming regions, such as $\rho$~Oph and Orion B.  The general shape
of the mass functions appears to be similar, but their intrinsic scales,
represented by their peak masses (for example), differ significantly.  The 
issue of the similarity in the shape of the mass function among 
star-forming regions will be discussed at length in Paper~III.

\end{enumerate}

\begin{acknowledgements}

M.~A.~R. has been supported by an Ontario Graduate Scholarship in Science
and Technology.  Both M.~A.~R. and C.~D.~W. are supported by the Natural
Sciences and Engineering Research Council of Canada (NSERC).  M.~A.~R.  
would like to thank E. Feigelson and N. Nedialkov for their helpful advice
on the statistical aspects of this work.  The James Clerk Maxwell
Telescope is operated by The Joint Astronomy Centre on behalf of the
Particle Physics and Astronomy Research Council of the United Kingdom, the
Netherlands Organisation for Scientific Research, and the National
Research Council of Canada.

\end{acknowledgements}


\clearpage 

\clearpage

\clearpage

\clearpage 

\clearpage

\begin{deluxetable}{ccccccccccc}
\tabletypesize{\tiny}
\tablecaption{Properties of the \eight\ Clumps\label{tab:m17850clumps}}
\tablewidth{0pt}
\tablehead{
\colhead{Name} & \colhead{R.A.}  & \colhead{Dec.} & \colhead{R$_{\rm 
eff}$} &
\colhead{$S_{\rm peak}$\tablenotemark{a}} & \colhead{$S^{\rm 
int}_{850}$\tablenotemark{a}} &
\colhead{$\langle{\alpha}\rangle$\tablenotemark{b}} &
\colhead{$\langle {\rm T}_{d}\rangle$\tablenotemark{c}} &
\colhead{$M_{{\rm T}_{d}=30K}$\tablenotemark{a}} &
\colhead{$M_{{\rm T}_{d}}$\tablenotemark{a}} & \colhead{n$_{\rm psc}$\tablenotemark{d}} \\
\colhead{(M17-)} & \colhead{(J2000)} & \colhead{(J2000)} & 
\colhead{(pc)} & \colhead{(Jy beam$^{-1}$)} &
\colhead{(Jy)} & \colhead{} &
\colhead{(K)} & \colhead{($M_{\odot}$)} & \colhead{($M_{\odot}$)} & 
\colhead{}}
\startdata
                     SMM1 & 18 20.0 00.6 & -16 00 18 &  0.12 &   0.26$\pm$0.01 &   0.45$\pm$0.05 &            2.89 &              23 &     7.4$\pm$0.8 &        11$\pm$1 &   0\\
                     SMM2 & 18 20.0 00.9 & -16 02 00 &  0.02 &   0.12$\pm$0.01 &   0.07$\pm$0.01 &            1.91 &              10 &   1.13$\pm$0.07 &     6.4$\pm$0.4 &   0\\
                     SMM3 & 18 20.0 02.7 & -16 02 14 &  0.06 &   0.15$\pm$0.01 &   0.14$\pm$0.01 &            3.00 &              27 &     2.2$\pm$0.2 &     2.6$\pm$0.2 &   0\\
                     SMM4 & 18 20.0 03.0 & -16 00 44 &  0.13 &   0.32$\pm$0.01 &   0.73$\pm$0.08 &            2.38 &              14 &        12$\pm$1 &        40$\pm$4 &   0\\
                     SMM5 & 18 20.0 03.7 & -16 01 56 &  0.06 &   0.16$\pm$0.01 &   0.17$\pm$0.01 &            3.10 &              33 &     2.9$\pm$0.2 &     2.5$\pm$0.2 &   0\\
                     SMM6 & 18 20.0 04.7 & -16 01 10 &  0.06 &   0.29$\pm$0.01 &   0.32$\pm$0.03 &            2.99 &              27 &     5.4$\pm$0.5 &     6.3$\pm$0.5 &   0\\
                     SMM7 & 18 20.0 04.8 & -16 02 22 &  0.14 &   0.46$\pm$0.02 &     1.1$\pm$0.1 &            3.04 &              30 &        18$\pm$2 &        19$\pm$2 &   0\\
                     SMM8 & 18 20.0 05.4 & -16 01 28 &  0.04 &   0.55$\pm$0.02 &   0.66$\pm$0.06 &            3.35 &              85 &    10.9$\pm$0.9 &     3.2$\pm$0.2 &   0\\
                     SMM9 & 18 20.0 05.9 & -16 04 46 &  0.18 &   0.82$\pm$0.03 &     2.5$\pm$0.3 &            2.85 &              21 &        42$\pm$6 &        66$\pm$9 &   0\\
                    SMM10 & 18 20.0 06.1 & -16 04 20 &  0.10 &   0.64$\pm$0.03 &     1.1$\pm$0.1 &            2.73 &              19 &        19$\pm$2 &        37$\pm$4 &   0\\
                    SMM11 & 18 20.0 06.2 & -16 01 56 &  0.13 &   0.55$\pm$0.02 &     1.2$\pm$0.1 &            3.28 &              58 &        19$\pm$2 &         9$\pm$1 &   0\\
                    SMM12 & 18 20.0 06.3 & -16 03 10 &  0.05 &   0.12$\pm$0.01 &   0.09$\pm$0.01 &            2.75 &              19 &     1.5$\pm$0.1 &     2.8$\pm$0.2 &   0\\
                    SMM13 & 18 20.0 06.9 & -16 03 42 &  0.11 &   0.15$\pm$0.01 &   0.30$\pm$0.03 &            2.67 &              17 &     5.0$\pm$0.5 &    10.8$\pm$0.9 &   0\\
                    SMM14 & 18 20.0 07.0 & -16 05 36 &  0.09 &   0.18$\pm$0.01 &   0.24$\pm$0.02 &            1.54 &               9 &     4.0$\pm$0.3 &        33$\pm$3 &   0\\
                    SMM15 & 18 20.0 07.2 & -16 05 58 &  0.05 &   0.21$\pm$0.01 &   0.22$\pm$0.02 &            3.09 &              32 &     3.6$\pm$0.3 &     3.3$\pm$0.2 &   0\\
                    SMM16 & 18 20.0 08.0 & -16 06 14 &  0.06 &   0.21$\pm$0.01 &   0.21$\pm$0.02 &            2.67 &              18 &     3.4$\pm$0.3 &     7.3$\pm$0.5 &   0\\
                    SMM17 & 18 20.0 09.2 & -16 04 18 &  0.10 &   0.34$\pm$0.01 &   0.54$\pm$0.06 &            2.45 &              14 &     8.9$\pm$0.9 &        27$\pm$3 &   0\\
                    SMM18 & 18 20.0 09.4 & -16 03 20 &  0.21 &   0.55$\pm$0.02 &     2.0$\pm$0.3 &            3.30 &              63 &        33$\pm$5 &        13$\pm$2 &   0\\
                    SMM19 & 18 20.0 10.5 & -16 00 24 &  0.05 &   0.15$\pm$0.01 &   0.10$\pm$0.01 &            3.42 &             144 &     1.7$\pm$0.1 &   0.29$\pm$0.02 &   0\\
                    SMM20 & 18 20.0 10.6 & -16 01 50 &  0.15 &   0.28$\pm$0.01 &   0.70$\pm$0.08 &            3.88 &         \nodata &        12$\pm$1 &         \nodata &   0\\
                    SMM21 & 18 20.0 10.6 & -16 02 22 &  0.15 &   0.19$\pm$0.01 &   0.53$\pm$0.06 &            3.86 &         \nodata &         9$\pm$1 &         \nodata &   0\\
                    SMM22 & 18 20.0 10.8 & -15 58 50 &  0.18 &   0.38$\pm$0.02 &     0.9$\pm$0.1 &            2.16 &              12 &        14$\pm$2 &        62$\pm$8 &   0\\
                    SMM23 & 18 20.0 11.3 & -16 07 06 &  0.19 &   1.14$\pm$0.05 &     4.1$\pm$0.6 &            2.79 &              20 &        67$\pm$9 &      120$\pm$20 &   1\\
                    SMM24 & 18 20.0 11.3 & -16 06 32 &  0.16 &   0.51$\pm$0.02 &     1.7$\pm$0.2 &            2.67 &              18 &        28$\pm$3 &        60$\pm$5 &   1\\
                    SMM25 & 18 20.0 12.3 & -16 04 26 &  0.04 &   0.14$\pm$0.01 &   0.11$\pm$0.01 &            3.04 &              29 &     1.8$\pm$0.1 &     1.8$\pm$0.1 &   0\\
                    SMM26 & 18 20.0 12.9 & -16 04 10 &  0.07 &   0.16$\pm$0.01 &   0.18$\pm$0.01 &            3.44 &             220 &     2.9$\pm$0.2 &   0.31$\pm$0.02 &   0\\
                    SMM27 & 18 20.0 13.0 & -16 00 22 &  0.10 &   0.32$\pm$0.01 &   0.48$\pm$0.05 &            3.23 &              47 &     7.9$\pm$0.7 &     4.5$\pm$0.4 &   0\\
                    SMM28 & 18 20.0 13.1 & -16 04 46 &  0.07 &   0.21$\pm$0.01 &   0.27$\pm$0.02 &            2.84 &              21 &     4.5$\pm$0.4 &     7.2$\pm$0.6 &   0\\
                    SMM29 & 18 20.0 13.8 & -16 05 06 &  0.11 &   0.28$\pm$0.01 &   0.54$\pm$0.06 &            2.89 &              23 &     9.0$\pm$0.9 &        13$\pm$1 &   0\\
                    SMM30 & 18 20.0 14.5 & -16 06 14 &  0.17 &   0.66$\pm$0.03 &     2.0$\pm$0.3 &            3.06 &              30 &        34$\pm$4 &        34$\pm$4 &   0\\
                    SMM31 & 18 20.0 14.7 & -16 07 04 &  0.14 &   1.30$\pm$0.05 &     3.7$\pm$0.5 &            3.17 &              40 &        61$\pm$8 &        43$\pm$5 &   0\\
                    SMM32 & 18 20.0 14.9 & -16 07 54 &  0.14 &   2.19$\pm$0.09 &     4.8$\pm$0.6 &            3.58 &         \nodata &       80$\pm$10 &         \nodata &   0\\
                    SMM33 & 18 20.0 15.1 & -16 07 34 &  0.11 &   1.13$\pm$0.05 &     2.9$\pm$0.3 &            3.31 &              66 &        48$\pm$5 &        19$\pm$2 &   0\\
                    SMM34 & 18 20.0 15.2 & -16 04 46 &  0.09 &   0.33$\pm$0.01 &   0.43$\pm$0.04 &            2.77 &              20 &     7.1$\pm$0.7 &        13$\pm$1 &   0\\
                    SMM35 & 18 20.0 15.9 & -16 09 28 &  0.06 &   0.94$\pm$0.04 &     1.5$\pm$0.1 &            3.65 &         \nodata &        25$\pm$2 &         \nodata &   0\\
                    SMM36 & 18 20.0 16.0 & -16 08 54 &  0.15 &     3.0$\pm$0.1 &     6.9$\pm$0.9 &            3.59 &         \nodata &      110$\pm$20 &         \nodata &   0\\
                    SMM37 & 18 20.0 16.0 & -16 05 46 &  0.13 &   1.00$\pm$0.04 &     2.5$\pm$0.3 &            3.18 &              40 &        42$\pm$5 &        29$\pm$3 &   0\\
                    SMM38 & 18 20.0 16.5 & -16 02 22 &  0.13 &   0.20$\pm$0.01 &   0.42$\pm$0.04 &            3.31 &              67 &     7.0$\pm$0.7 &     2.7$\pm$0.3 &   0\\
                    SMM39 & 18 20.0 16.7 & -16 07 48 &  0.06 &   0.31$\pm$0.01 &   0.41$\pm$0.03 &            2.61 &              16 &     6.8$\pm$0.6 &        16$\pm$1 &   0\\
                    SMM40 & 18 20.0 16.7 & -16 01 54 &  0.12 &   0.14$\pm$0.01 &   0.23$\pm$0.02 &            3.10 &              33 &     3.9$\pm$0.4 &     3.4$\pm$0.3 &   0\\
                    SMM41 & 18 20.0 16.9 & -16 05 20 &  0.13 &   0.46$\pm$0.02 &     1.1$\pm$0.1 &            2.87 &              22 &        18$\pm$2 &        27$\pm$3 &   0\\
                    SMM42 & 18 20.0 17.2 & -16 00 50 &  0.12 &   0.17$\pm$0.01 &   0.30$\pm$0.03 &            2.57 &              16 &     5.0$\pm$0.5 &        13$\pm$1 &   0\\
                    SMM43 & 18 20.0 17.9 & -16 08 32 &  0.05 &   0.34$\pm$0.01 &   0.20$\pm$0.01 &            2.16 &              12 &     3.3$\pm$0.2 &    14.3$\pm$0.8 &   0\\
                    SMM44 & 18 20.0 18.1 & -16 04 22 &  0.09 &   0.36$\pm$0.01 &   0.38$\pm$0.04 &            3.62 &         \nodata &     6.3$\pm$0.6 &         \nodata &   0\\
                    SMM45 & 18 20.0 18.7 & -16 06 10 &  0.11 &   0.40$\pm$0.02 &   0.78$\pm$0.08 &            2.39 &              14 &        13$\pm$1 &        42$\pm$4 &   0\\
   SMM46\tablenotemark{e} & 18 20.0 19.5 & -16 09 40 &  0.06 &   1.86$\pm$0.07 &     2.5$\pm$0.3 &            3.43 &             164 &        42$\pm$5 &     6.1$\pm$0.6 &   0\\
                    SMM47 & 18 20.0 20.2 & -16 07 16 &  0.03 &   0.14$\pm$0.01 &   0.05$\pm$0.01 &         \nodata &         \nodata &   0.81$\pm$0.04 &         \nodata &   0\\
   SMM48\tablenotemark{e} & 18 20.0 22.0 & -16 06 14 &  0.10 &   0.46$\pm$0.02 &   0.67$\pm$0.07 &            3.22 &              46 &        11$\pm$1 &     6.6$\pm$0.7 &   0\\
                    SMM49 & 18 20.0 24.1 & -15 58 36 &  0.14 &   0.20$\pm$0.01 &   0.54$\pm$0.06 &            3.73 &         \nodata &         9$\pm$1 &         \nodata &   0\\
                    SMM50 & 18 20.0 24.7 & -16 06 00 &  0.07 &   0.53$\pm$0.02 &   0.85$\pm$0.08 &            3.45 &             235 &        14$\pm$1 &     1.4$\pm$0.1 &   0\\
                    SMM51 & 18 20.0 24.7 & -15 59 18 &  0.06 &   0.38$\pm$0.02 &   0.59$\pm$0.05 &            3.31 &              65 &     9.8$\pm$0.8 &     3.9$\pm$0.3 &   1\\
                    SMM52 & 18 20.0 25.1 & -16 03 48 &  0.03 &   0.59$\pm$0.02 &   0.55$\pm$0.05 &            3.58 &         \nodata &     9.1$\pm$0.8 &         \nodata &   0\\
                    SMM53 & 18 20.0 25.1 & -15 59 08 &  0.08 &   0.29$\pm$0.01 &   0.45$\pm$0.04 &            3.38 &             105 &     7.6$\pm$0.6 &     1.8$\pm$0.1 &   0\\
                    SMM54 & 18 20.0 25.4 & -16 06 18 &  0.11 &   0.55$\pm$0.02 &   0.87$\pm$0.09 &            3.58 &         \nodata &        15$\pm$2 &         \nodata &   0\\
   SMM55\tablenotemark{e} & 18 20.0 25.5 & -16 05 28 &  0.12 &   0.49$\pm$0.02 &     1.0$\pm$0.1 &            3.48 &         \nodata &        17$\pm$2 &         \nodata &   0\\
                    SMM56 & 18 20.0 25.5 & -16 05 54 &  0.05 &   0.50$\pm$0.02 &   0.82$\pm$0.07 &            3.40 &             126 &        14$\pm$1 &     2.6$\pm$0.2 &   0\\
                    SMM57 & 18 20.0 25.9 & -16 02 32 &  0.09 &   0.32$\pm$0.01 &   0.46$\pm$0.04 &            2.71 &              18 &     7.6$\pm$0.7 &        15$\pm$1 &   0\\
                    SMM58 & 18 20.0 25.9 & -15 59 34 &  0.12 &   0.32$\pm$0.01 &   0.66$\pm$0.07 &            3.03 &              29 &        11$\pm$1 &        12$\pm$1 &   0\\
                    SMM59 & 18 20.0 26.2 & -16 03 04 &  0.02 &   0.13$\pm$0.01 &   0.07$\pm$0.01 &         \nodata &         \nodata &   1.17$\pm$0.07 &         \nodata &   0\\
                    SMM60 & 18 20.0 26.5 & -16 02 14 &  0.11 &   0.33$\pm$0.01 &   0.71$\pm$0.07 &            2.87 &              22 &        12$\pm$1 &        18$\pm$2 &   0\\
   SMM61\tablenotemark{e} & 18 20.0 26.6 & -16 07 38 &  0.14 &   0.34$\pm$0.01 &     1.0$\pm$0.2 &            2.27 &              13 &        17$\pm$3 &       70$\pm$10 &   0\\
                    SMM62 & 18 20.0 26.9 & -16 01 34 &  0.14 &   0.36$\pm$0.01 &     1.1$\pm$0.1 &            2.18 &              12 &        18$\pm$2 &        74$\pm$8 &   0\\
                    SMM63 & 18 20.0 27.0 & -16 00 48 &  0.12 &   0.25$\pm$0.01 &   0.52$\pm$0.05 &            1.69 &               9 &     8.6$\pm$0.9 &        61$\pm$6 &   0\\
   SMM64\tablenotemark{e} & 18 20.0 27.3 & -16 08 26 &  0.12 &   0.73$\pm$0.03 &     1.3$\pm$0.2 &            2.72 &              18 &        21$\pm$3 &        41$\pm$5 &   0\\
   SMM65\tablenotemark{e} & 18 20.0 27.3 & -16 07 12 &  0.13 &   0.58$\pm$0.02 &     1.3$\pm$0.2 &            1.46 &               8 &        22$\pm$3 &      200$\pm$20 &   1\\
                    SMM66 & 18 20.0 27.6 & -16 00 14 &  0.12 &   0.40$\pm$0.02 &   0.82$\pm$0.09 &            2.77 &              19 &        14$\pm$2 &        25$\pm$3 &   0\\
                    SMM67 & 18 20.0 27.9 & -16 00 34 &  0.13 &   0.42$\pm$0.02 &     1.2$\pm$0.1 &            2.19 &              12 &        20$\pm$2 &        83$\pm$9 &   1\\
   SMM68\tablenotemark{e} & 18 20.0 28.0 & -16 08 58 &  0.12 &   0.26$\pm$0.01 &     0.5$\pm$0.1 &            0.45 &               6 &         9$\pm$2 &      180$\pm$30 &   2\\
                    SMM69 & 18 20.0 28.7 & -16 05 36 &  0.10 &   0.19$\pm$0.01 &   0.28$\pm$0.03 &         \nodata &         \nodata &     4.7$\pm$0.4 &         \nodata &   0\\
                    SMM70 & 18 20.0 28.8 & -16 01 28 &  0.16 &   1.07$\pm$0.04 &     4.2$\pm$0.6 &            2.61 &              16 &        69$\pm$9 &      170$\pm$20 &   1\\
   SMM71\tablenotemark{e} & 18 20.0 28.8 & -16 07 28 &  0.17 &   0.50$\pm$0.02 &     1.4$\pm$0.3 &            2.55 &              16 &        23$\pm$4 &       60$\pm$10 &   0\\
   SMM72\tablenotemark{e} & 18 20.0 29.5 & -16 09 20 &  0.08 &   0.18$\pm$0.01 &   0.18$\pm$0.03 &         \nodata &         \nodata &     2.9$\pm$0.5 &         \nodata &   0\\
                    SMM73 & 18 20.0 29.9 & -16 02 04 &  0.14 &   1.49$\pm$0.06 &     4.4$\pm$0.6 &            2.73 &              19 &        73$\pm$9 &      140$\pm$20 &   2\\
   SMM74\tablenotemark{e} & 18 20.0 29.9 & -16 09 32 &  0.04 &   0.25$\pm$0.01 &   0.18$\pm$0.02 &         \nodata &         \nodata &     3.0$\pm$0.4 &         \nodata &   0\\
                    SMM75 & 18 20.0 30.2 & -16 05 16 &  0.10 &   0.25$\pm$0.01 &   0.35$\pm$0.03 &            1.66 &               9 &     5.8$\pm$0.5 &        43$\pm$4 &   0\\
                    SMM76 & 18 20.0 30.5 & -16 04 12 &  0.08 &   0.15$\pm$0.01 &   0.20$\pm$0.02 &         \nodata &         \nodata &     3.3$\pm$0.3 &         \nodata &   0\\
   SMM77\tablenotemark{e} & 18 20.0 30.6 & -16 08 40 &  0.12 &   1.87$\pm$0.07 &     3.9$\pm$0.5 &            2.44 &              14 &        65$\pm$9 &      200$\pm$30 &   0\\
                    SMM78 & 18 20.0 31.0 & -15 57 48 &  0.03 &   0.16$\pm$0.01 &   0.09$\pm$0.01 &         \nodata &         \nodata &     1.4$\pm$0.1 &         \nodata &   0\\
                    SMM79 & 18 20.0 31.3 & -15 58 04 &  0.12 &   0.17$\pm$0.01 &   0.32$\pm$0.03 &         \nodata &         \nodata &     5.4$\pm$0.6 &         \nodata &   0\\
                    SMM80 & 18 20.0 31.6 & -16 02 16 &  0.14 &   0.79$\pm$0.03 &     2.0$\pm$0.3 &            2.69 &              18 &        34$\pm$4 &        70$\pm$8 &   0\\
   SMM81\tablenotemark{e} & 18 20.0 31.7 & -16 08 28 &  0.16 &   1.29$\pm$0.05 &     4.1$\pm$0.7 &            2.68 &              18 &       70$\pm$10 &      140$\pm$20 &   0\\
                    SMM82 & 18 20.0 32.2 & -16 01 00 &  0.12 &   0.85$\pm$0.03 &     2.8$\pm$0.3 &            2.67 &              18 &        46$\pm$6 &      100$\pm$10 &   0\\
                    SMM83 & 18 20.0 32.2 & -16 09 24 &  0.05 &   0.20$\pm$0.01 &   0.16$\pm$0.01 &         \nodata &         \nodata &     2.7$\pm$0.2 &         \nodata &   1\\
                    SMM84 & 18 20.0 32.6 & -16 00 34 &  0.12 &   0.53$\pm$0.02 &     1.3$\pm$0.1 &            2.23 &              12 &        22$\pm$2 &        87$\pm$9 &   0\\
                    SMM85 & 18 20.0 32.7 & -16 02 48 &  0.11 &   0.24$\pm$0.01 &   0.38$\pm$0.04 &            3.27 &              54 &     6.3$\pm$0.7 &     3.0$\pm$0.3 &   0\\
   SMM86\tablenotemark{e} & 18 20.0 32.9 & -16 01 44 &  0.14 &     3.7$\pm$0.1 &         7$\pm$1 &            3.14 &              37 &      120$\pm$20 &       90$\pm$20 &   1\\
                    SMM87 & 18 20.0 32.9 & -15 58 58 &  0.18 &   0.99$\pm$0.04 &     3.3$\pm$0.5 &            2.88 &              23 &        55$\pm$8 &       80$\pm$10 &   0\\
                    SMM88 & 18 20.0 32.9 & -16 04 08 &  0.18 &   0.83$\pm$0.03 &     2.3$\pm$0.3 &            3.19 &              42 &        38$\pm$5 &        25$\pm$3 &   0\\
                    SMM89 & 18 20.0 32.9 & -16 07 16 &  0.16 &   0.52$\pm$0.02 &     1.4$\pm$0.2 &            2.86 &              22 &        23$\pm$3 &        35$\pm$4 &   0\\
                    SMM90 & 18 20.0 33.8 & -16 01 12 &  0.11 &   1.55$\pm$0.06 &     4.2$\pm$0.5 &            2.90 &              23 &        69$\pm$8 &      100$\pm$10 &   0\\
                    SMM91 & 18 20.0 34.0 & -15 58 06 &  0.17 &   0.65$\pm$0.03 &     2.2$\pm$0.3 &            2.76 &              19 &        37$\pm$5 &        69$\pm$9 &   0\\
                    SMM92 & 18 20.0 34.1 & -15 59 56 &  0.20 &   1.20$\pm$0.05 &     5.8$\pm$0.8 &            2.78 &              20 &      100$\pm$10 &      170$\pm$20 &   0\\
   SMM93\tablenotemark{e} & 18 20.0 34.1 & -16 07 46 &  0.17 &   0.62$\pm$0.02 &     2.5$\pm$0.3 &            2.60 &              16 &        41$\pm$6 &      100$\pm$10 &   0\\
                    SMM94 & 18 20.0 34.2 & -15 58 38 &  0.14 &   0.67$\pm$0.03 &     1.5$\pm$0.2 &            2.82 &              21 &        26$\pm$3 &        43$\pm$5 &   0\\
   SMM95\tablenotemark{e} & 18 20.0 35.1 & -16 08 28 &  0.19 &   1.41$\pm$0.06 &     5.0$\pm$0.8 &            2.67 &              17 &       80$\pm$10 &      180$\pm$30 &   0\\
                    SMM96 & 18 20.0 35.1 & -16 04 54 &  0.23 &   0.98$\pm$0.04 &     4.4$\pm$0.6 &            3.05 &              30 &       70$\pm$10 &       70$\pm$10 &   0\\
                    SMM97 & 18 20.0 35.2 & -16 00 44 &  0.21 &   1.28$\pm$0.05 &     6.5$\pm$0.9 &            2.79 &              20 &      110$\pm$20 &      190$\pm$30 &   1\\
                    SMM98 & 18 20.0 35.4 & -16 06 04 &  0.22 &   1.55$\pm$0.06 &     5.5$\pm$0.8 &            3.26 &              53 &       90$\pm$10 &        45$\pm$6 &   1\\
                    SMM99 & 18 20.0 37.2 & -15 57 56 &  0.10 &   0.60$\pm$0.02 &     1.0$\pm$0.1 &            2.65 &              17 &        16$\pm$2 &        36$\pm$3 &   0\\
                   SMM100 & 18 20.0 37.3 & -15 59 38 &  0.20 &   0.91$\pm$0.04 &     3.9$\pm$0.5 &            2.49 &              15 &        65$\pm$9 &      180$\pm$20 &   0\\
                   SMM101 & 18 20.0 37.7 & -16 01 50 &  0.09 &   0.93$\pm$0.04 &     1.4$\pm$0.1 &            2.62 &              17 &        23$\pm$2 &        55$\pm$5 &   0\\
                   SMM102 & 18 20.0 37.9 & -15 57 36 &  0.06 &   0.51$\pm$0.02 &   0.68$\pm$0.06 &            2.87 &              22 &        11$\pm$1 &        17$\pm$1 &   0\\
                   SMM103 & 18 20.0 37.9 & -15 57 16 &  0.01 &   0.40$\pm$0.02 &   0.40$\pm$0.03 &            2.93 &              24 &     6.7$\pm$0.5 &     9.0$\pm$0.6 &   0\\
                   SMM104 & 18 20.0 37.9 & -16 02 56 &  0.12 &   0.29$\pm$0.01 &   0.63$\pm$0.07 &            2.87 &              22 &        10$\pm$1 &        16$\pm$2 &   0\\
                   SMM105 & 18 20.0 37.9 & -16 06 22 &  0.07 &   0.20$\pm$0.01 &   0.23$\pm$0.02 &            2.28 &              13 &     3.9$\pm$0.3 &        14$\pm$1 &   0\\
                   SMM106 & 18 20.0 38.0 & -16 01 34 &  0.18 &   1.16$\pm$0.05 &     3.0$\pm$0.4 &            2.82 &              21 &        50$\pm$7 &       80$\pm$10 &   0\\
                   SMM107 & 18 20.0 38.0 & -16 03 20 &  0.05 &   0.21$\pm$0.01 &   0.22$\pm$0.02 &            2.85 &              22 &     3.7$\pm$0.3 &     5.8$\pm$0.4 &   0\\
                   SMM108 & 18 20.0 38.1 & -15 59 16 &  0.16 &   0.99$\pm$0.04 &     2.4$\pm$0.3 &            2.72 &              18 &        40$\pm$5 &       80$\pm$10 &   1\\
                   SMM109 & 18 20.0 38.5 & -15 58 08 &  0.13 &   0.58$\pm$0.02 &     1.1$\pm$0.1 &            2.74 &              19 &        19$\pm$2 &        36$\pm$4 &   0\\
  SMM110\tablenotemark{e} & 18 20.0 38.7 & -16 09 04 &  0.14 &   0.26$\pm$0.01 &     0.5$\pm$0.1 &         \nodata &         \nodata &         9$\pm$2 &         \nodata &   0\\
                   SMM111 & 18 20.0 39.0 & -16 07 52 &  0.19 &   0.40$\pm$0.02 &     1.6$\pm$0.2 &            1.99 &              11 &        27$\pm$4 &      140$\pm$20 &   0\\
                   SMM112 & 18 20.0 39.1 & -16 06 00 &  0.16 &   0.55$\pm$0.02 &     1.8$\pm$0.2 &            2.60 &              16 &        30$\pm$4 &        71$\pm$9 &   1\\
                   SMM113 & 18 20.0 39.5 & -16 06 52 &  0.10 &   0.43$\pm$0.02 &   0.74$\pm$0.08 &            3.06 &              30 &        12$\pm$1 &        12$\pm$1 &   0\\
                   SMM114 & 18 20.0 39.7 & -16 02 16 &  0.14 &   0.47$\pm$0.02 &     1.5$\pm$0.2 &            1.80 &              10 &        25$\pm$3 &      160$\pm$20 &   1\\
  SMM115\tablenotemark{e} & 18 20.0 39.7 & -16 09 22 &  0.10 &   0.26$\pm$0.01 &   0.38$\pm$0.06 &         \nodata &         \nodata &     6.2$\pm$0.9 &         \nodata &   0\\
                   SMM116 & 18 20.0 40.6 & -16 07 04 &  0.08 &   0.30$\pm$0.01 &   0.35$\pm$0.03 &            2.90 &              23 &     5.9$\pm$0.5 &     8.3$\pm$0.6 &   0\\
                   SMM117 & 18 20.0 41.2 & -16 02 22 &  0.12 &   0.21$\pm$0.01 &   0.40$\pm$0.04 &         \nodata &         \nodata &     6.6$\pm$0.7 &         \nodata &   0\\
  SMM118\tablenotemark{e} & 18 20.0 41.6 & -16 08 26 &  0.12 &   0.59$\pm$0.02 &     1.1$\pm$0.2 &            2.64 &              17 &        19$\pm$3 &        43$\pm$6 &   0\\
                   SMM119 & 18 20.0 42.0 & -16 07 22 &  0.11 &   0.40$\pm$0.02 &   0.59$\pm$0.06 &            2.10 &              11 &        10$\pm$1 &        45$\pm$4 &   0\\
                   SMM120 & 18 20.0 42.7 & -16 01 36 &  0.03 &   0.16$\pm$0.01 &   0.10$\pm$0.01 &         \nodata &         \nodata &   1.64$\pm$0.09 &         \nodata &   0\\
  SMM121\tablenotemark{e} & 18 20.0 42.9 & -16 08 26 &  0.19 &   0.59$\pm$0.02 &     1.8$\pm$0.3 &            2.82 &              21 &        30$\pm$5 &        50$\pm$7 &   1\\
\enddata
\tablenotetext{a}{The uncertainties stated in this table are composed of
the uncertainties in the gain calibration, the sky opacities, and the
corrections due to the error beam.  These uncertainties are typically
significantly larger than the random measurement errors associated with
the rms flux of the map.  The exception is the peak flux, where the random
error of $\sigma = 0.027$ Jy beam$^{-1}$ dominates the systematic error
for the clumps with lower peak fluxes.}
\tablenotetext{b}{The systematic uncertainty in the spectral index,
$\alpha$, is 13\%.  These systematic uncertainties, which are composed of 
the uncertainties in the 
gain calibration and sky opacities, dominate the random errors on the 
spectral index.}
\tablenotetext{c}{See Section \ref{sec:alphaandt} for a discussion of the
uncertainties in the temperatures.  Temperatures are omitted where high
spectral index makes them incalculable, or where no reliable spectral
index can be calculated (see text).  All of the temperatures above 40~K 
should be considered highly uncertain, indicating only that a clump is 
probably hot.} 
\tablenotetext{d}{Number of MSX point sources contained 
within the clump's 0.5$S_{\rm peak}$ contour.}
\tablenotetext{e}{Denotes a clump to which corrections for free-free
emission have been applied in the calculation of the spectral index, dust
temperature, and masses.  The free-free correction has \emph{not} been
applied to the peak and integrated fluxes listed here.}
\end{deluxetable}

\clearpage

\begin{deluxetable}{cccccccc}
\tabletypesize{\scriptsize}
\tablecaption{Properties of the \four\ Clumps\label{tab:m17450clumps}}
\tablewidth{0pt}
\tablehead{
\colhead{Name\tablenotemark{a}} & \colhead{R.A.}  & \colhead{Dec.} &
\colhead{R$_{\rm eff}$} &
\colhead{$S_{\rm peak}$\tablenotemark{b}} & \colhead{$S^{\rm
int}_{450}$\tablenotemark{b}} &
\colhead{$M_{{\rm T}_{d}=30K}$\tablenotemark{b}} & \colhead{n$_{\rm psc}$\tablenotemark{c}} \\
\colhead{(M17-)} & \colhead{(J2000)} & \colhead{(J2000)} &
\colhead{(pc)} & \colhead{Jy beam$^{-1}$} & \colhead{(Jy)} & \colhead{($M_{\odot}$)} & \colhead{} }
\startdata
                    SMM4A & 18 20.0 03.0 & -16 00 46 &  0.06 &     1.5$\pm$0.2 &     2.4$\pm$0.3 &     4.2$\pm$0.6 &   0 \\
                    SMM7B & 18 20.0 03.6 & -16 02 22 &  0.07 &     1.6$\pm$0.2 &     2.6$\pm$0.4 &     4.5$\pm$0.7 &   0 \\
                    SMM7A & 18 20.0 04.9 & -16 02 24 &  0.09 &     2.4$\pm$0.3 &     5.6$\pm$0.9 &        10$\pm$2 &   0 \\
                    SMM8A & 18 20.0 05.4 & -16 01 30 &  0.09 &     3.7$\pm$0.4 &         8$\pm$1 &        13$\pm$2 &   0 \\
                    SMM9B & 18 20.0 05.9 & -16 04 44 &  0.11 &     3.4$\pm$0.4 &        11$\pm$2 &        19$\pm$3 &   0 \\
                   SMM11B & 18 20.0 06.1 & -16 01 52 &  0.11 &     3.3$\pm$0.4 &         8$\pm$1 &        14$\pm$3 &   0 \\
                   SMM10A & 18 20.0 06.1 & -16 04 22 &  0.09 &     2.9$\pm$0.3 &         7$\pm$1 &        13$\pm$2 &   0 \\
                   SMM15A & 18 20.0 06.7 & -16 05 54 &  0.06 &     1.3$\pm$0.2 &     2.0$\pm$0.3 &     3.4$\pm$0.5 &   0 \\
                    SMM9A & 18 20.0 07.4 & -16 04 52 &  0.09 &     2.4$\pm$0.3 &         6$\pm$1 &        10$\pm$2 &   0 \\
                   SMM11A & 18 20.0 07.6 & -16 02 30 &  0.07 &     1.8$\pm$0.2 &     2.9$\pm$0.4 &     5.1$\pm$0.8 &   0 \\
                   SMM18A & 18 20.0 09.0 & -16 03 50 &  0.06 &     1.3$\pm$0.2 &     2.4$\pm$0.3 &     4.2$\pm$0.6 &   0 \\
                   SMM18B & 18 20.0 09.1 & -16 03 24 &  0.15 &     2.7$\pm$0.3 &        13$\pm$2 &        22$\pm$4 &   0 \\
                   SMM17A & 18 20.0 09.2 & -16 04 18 &  0.05 &     1.6$\pm$0.2 &     1.9$\pm$0.3 &     3.3$\pm$0.5 &   0 \\
                   SMM18C & 18 20.0 09.4 & -16 02 44 &  0.06 &     1.8$\pm$0.2 &     2.8$\pm$0.4 &     4.8$\pm$0.7 &   0 \\
                   SMM20B & 18 20.0 09.5 & -16 01 22 &  0.07 &     2.2$\pm$0.3 &     2.8$\pm$0.4 &     4.9$\pm$0.7 &   0 \\
                   SMM23B & 18 20.0 10.1 & -16 07 08 &  0.11 &     4.2$\pm$0.5 &        12$\pm$2 &        21$\pm$4 &   0 \\
                   SMM22A & 18 20.0 10.5 & -15 58 48 &  0.04 &     1.9$\pm$0.2 &     1.7$\pm$0.2 &     3.0$\pm$0.4 &   0 \\
                   SMM21B & 18 20.0 10.5 & -16 02 34 &  0.05 &     1.4$\pm$0.2 &     2.0$\pm$0.3 &     3.5$\pm$0.5 &   0 \\
                   SMM20A & 18 20.0 10.6 & -16 02 02 &  0.11 &     1.8$\pm$0.2 &         7$\pm$1 &        11$\pm$2 &   0 \\
                   SMM24A & 18 20.0 10.9 & -16 06 30 &  0.11 &     2.4$\pm$0.3 &         7$\pm$1 &        12$\pm$2 &   0 \\
                   SMM21A & 18 20.0 10.9 & -16 02 48 &  0.07 &     1.3$\pm$0.2 &     2.6$\pm$0.4 &     4.6$\pm$0.6 &   0 \\
                   SMM21C & 18 20.0 11.2 & -16 02 24 &  0.05 &     1.4$\pm$0.2 &     1.7$\pm$0.2 &     3.0$\pm$0.4 &   0 \\
                   SMM23A & 18 20.0 11.5 & -16 07 08 &  0.12 &     4.9$\pm$0.6 &        13$\pm$3 &        23$\pm$4 &   0 \\
                   SMM27A & 18 20.0 13.0 & -16 00 28 &  0.07 &     2.0$\pm$0.2 &     4.0$\pm$0.6 &         7$\pm$1 &   0 \\
                   SMM28A & 18 20.0 13.1 & -16 04 46 &  0.08 &     1.9$\pm$0.2 &     3.4$\pm$0.5 &     5.8$\pm$0.9 &   0 \\
                   SMM29A & 18 20.0 13.3 & -16 05 02 &  0.06 &     1.5$\pm$0.2 &     2.6$\pm$0.4 &     4.5$\pm$0.6 &   0 \\
                   SMM30A & 18 20.0 14.4 & -16 06 12 &  0.11 &     3.0$\pm$0.4 &        10$\pm$2 &        17$\pm$3 &   0 \\
                   SMM30B & 18 20.0 14.4 & -16 05 56 &  0.07 &     2.9$\pm$0.3 &     5.4$\pm$0.8 &         9$\pm$1 &   0 \\
                   SMM31A & 18 20.0 14.5 & -16 07 00 &  0.16 &     5.9$\pm$0.7 &        29$\pm$6 &       50$\pm$10 &   0 \\
                   SMM32A & 18 20.0 14.8 & -16 07 56 &  0.17 &        13$\pm$2 &       50$\pm$10 &       90$\pm$20 &   0 \\
                   SMM37B & 18 20.0 14.8 & -16 05 30 &  0.10 &     3.0$\pm$0.4 &         9$\pm$2 &        15$\pm$3 &   0 \\
                   SMM33A & 18 20.0 15.4 & -16 07 34 &  0.12 &     6.4$\pm$0.8 &        23$\pm$4 &        40$\pm$8 &   0 \\
                   SMM34A & 18 20.0 15.4 & -16 04 48 &  0.06 &     2.0$\pm$0.2 &     2.8$\pm$0.4 &     4.9$\pm$0.7 &   0 \\
                   SMM35A & 18 20.0 15.5 & -16 09 28 &  0.14 &     5.4$\pm$0.6 &        23$\pm$5 &        40$\pm$8 &   0 \\
                   SMM36A & 18 20.0 16.0 & -16 08 54 &  0.18 &        17$\pm$2 &       70$\pm$10 &      120$\pm$30 &   0 \\
                   SMM37A & 18 20.0 16.2 & -16 05 50 &  0.11 &     5.1$\pm$0.6 &        14$\pm$2 &        24$\pm$4 &   0 \\
                   SMM33B & 18 20.0 16.5 & -16 07 26 &  0.08 &     4.0$\pm$0.5 &         7$\pm$1 &        13$\pm$2 &   0 \\
                   SMM41A & 18 20.0 16.9 & -16 05 30 &  0.09 &     2.3$\pm$0.3 &     5.8$\pm$0.8 &        10$\pm$1 &   0 \\
                   SMM45A & 18 20.0 16.9 & -16 06 28 &  0.06 &     1.6$\pm$0.2 &     2.4$\pm$0.3 &     4.2$\pm$0.6 &   0 \\
                   SMM43A & 18 20.0 17.6 & -16 08 32 &  0.03 &     1.9$\pm$0.2 &     1.8$\pm$0.2 &     3.1$\pm$0.4 &   0 \\
                   SMM44A & 18 20.0 18.1 & -16 04 24 &  0.06 &     2.4$\pm$0.3 &     3.9$\pm$0.6 &         7$\pm$1 &   0 \\
                   SMM45B & 18 20.0 18.7 & -16 06 08 &  0.06 &     2.0$\pm$0.2 &     2.9$\pm$0.4 &     5.0$\pm$0.7 &   0 \\
                   SMM46A & 18 20.0 19.4 & -16 09 44 &  0.14 &        10$\pm$1 &        27$\pm$5 &        46$\pm$9 &   0 \\
                   SMM48A & 18 20.0 21.9 & -16 06 12 &  0.10 &     2.0$\pm$0.2 &         6$\pm$1 &        10$\pm$2 &   1 \\
                   SMM49A & 18 20.0 23.0 & -15 58 38 &  0.10 &     1.6$\pm$0.2 &     5.7$\pm$0.8 &        10$\pm$1 &   0 \\
                   SMM51A & 18 20.0 24.1 & -15 59 06 &  0.12 &     2.0$\pm$0.2 &         8$\pm$2 &        14$\pm$3 &   0 \\
                   SMM52A & 18 20.0 25.1 & -16 03 50 &  0.07 &     3.7$\pm$0.4 &         7$\pm$1 &        13$\pm$2 &   0 \\
                   SMM54A & 18 20.0 25.1 & -16 06 20 &  0.12 &     3.6$\pm$0.4 &        11$\pm$2 &        19$\pm$4 &   0 \\
                   SMM55A & 18 20.0 25.1 & -16 05 28 &  0.12 &     2.8$\pm$0.3 &        10$\pm$2 &        18$\pm$3 &   0 \\
                   SMM61A & 18 20.0 25.5 & -16 07 36 &  0.07 &     1.5$\pm$0.2 &     3.1$\pm$0.5 &     5.3$\pm$0.8 &   0 \\
                   SMM56A & 18 20.0 25.8 & -16 05 58 &  0.14 &     2.7$\pm$0.3 &        14$\pm$2 &        24$\pm$4 &   0 \\
                   SMM60A & 18 20.0 25.8 & -16 02 04 &  0.08 &     1.9$\pm$0.2 &     3.9$\pm$0.6 &         7$\pm$1 &   0 \\
                   SMM57A & 18 20.0 25.8 & -16 02 22 &  0.07 &     1.7$\pm$0.2 &     3.2$\pm$0.4 &     5.6$\pm$0.7 &   0 \\
                   SMM62A & 18 20.0 26.6 & -16 01 46 &  0.07 &     1.7$\pm$0.2 &     3.0$\pm$0.5 &     5.2$\pm$0.8 &   0 \\
                   SMM65A & 18 20.0 27.2 & -16 07 12 &  0.03 &     1.9$\pm$0.2 &     1.5$\pm$0.2 &     2.5$\pm$0.3 &   0 \\
                   SMM64A & 18 20.0 27.3 & -16 08 28 &  0.08 &     5.3$\pm$0.6 &         7$\pm$1 &        13$\pm$2 &   0 \\
                   SMM67A & 18 20.0 28.1 & -16 00 30 &  0.09 &     1.9$\pm$0.2 &     5.0$\pm$0.8 &         9$\pm$1 &   0 \\
                   SMM70A & 18 20.0 28.7 & -16 01 24 &  0.15 &     4.1$\pm$0.5 &        18$\pm$4 &        32$\pm$6 &   0 \\
                   SMM68A & 18 20.0 29.0 & -16 08 56 &  0.02 &     1.7$\pm$0.2 &     1.1$\pm$0.1 &     1.8$\pm$0.2 &   0 \\
                   SMM71A & 18 20.0 29.5 & -16 07 32 &  0.14 &     1.8$\pm$0.2 &        10$\pm$2 &        18$\pm$3 &   0 \\
                   SMM73A & 18 20.0 29.7 & -16 02 06 &  0.14 &     5.7$\pm$0.7 &        23$\pm$4 &        39$\pm$8 &   1 \\
                   SMM80A & 18 20.0 29.8 & -16 02 40 &  0.10 &     3.4$\pm$0.4 &         9$\pm$1 &        16$\pm$2 &   0 \\
                   SMM77A & 18 20.0 30.8 & -16 08 46 &  0.09 &     6.8$\pm$0.8 &        16$\pm$3 &        28$\pm$5 &   0 \\
                   SMM81B & 18 20.0 31.0 & -16 08 18 &  0.13 &     4.6$\pm$0.5 &        17$\pm$3 &        29$\pm$6 &   0 \\
                   SMM80B & 18 20.0 31.5 & -16 02 16 &  0.10 &     3.9$\pm$0.5 &         9$\pm$2 &        16$\pm$3 &   0 \\
                   SMM82A & 18 20.0 32.2 & -16 01 02 &  0.15 &     3.3$\pm$0.4 &        18$\pm$4 &        31$\pm$6 &   0 \\
                   SMM81A & 18 20.0 32.4 & -16 08 28 &  0.12 &     5.0$\pm$0.6 &        14$\pm$3 &        24$\pm$4 &   0 \\
                   SMM87A & 18 20.0 32.6 & -15 59 00 &  0.14 &     4.2$\pm$0.5 &        19$\pm$4 &        33$\pm$7 &   0 \\
                   SMM85A & 18 20.0 32.6 & -16 02 50 &  0.06 &     1.6$\pm$0.2 &     3.1$\pm$0.4 &     5.3$\pm$0.7 &   0 \\
                   SMM86A & 18 20.0 32.7 & -16 01 48 &  0.19 &        25$\pm$3 &       90$\pm$20 &      160$\pm$40 &   1 \\
                   SMM88A & 18 20.0 32.9 & -16 04 12 &  0.13 &     4.7$\pm$0.6 &        14$\pm$3 &        24$\pm$5 &   0 \\
                   SMM93A & 18 20.0 33.0 & -16 07 44 &  0.19 &     2.4$\pm$0.3 &        18$\pm$4 &        30$\pm$6 &   0 \\
                   SMM88B & 18 20.0 33.0 & -16 03 44 &  0.07 &     2.0$\pm$0.2 &     4.1$\pm$0.6 &         7$\pm$1 &   0 \\
                   SMM91A & 18 20.0 33.7 & -15 58 10 &  0.15 &     2.9$\pm$0.4 &        15$\pm$3 &        25$\pm$4 &   0 \\
                   SMM92B & 18 20.0 33.8 & -15 59 52 &  0.15 &     4.8$\pm$0.6 &        23$\pm$5 &        39$\pm$8 &   0 \\
                   SMM92A & 18 20.0 34.1 & -16 00 18 &  0.14 &     3.2$\pm$0.4 &        15$\pm$3 &        26$\pm$5 &   0 \\
                   SMM94A & 18 20.0 34.2 & -15 58 38 &  0.11 &     3.4$\pm$0.4 &        11$\pm$2 &        18$\pm$3 &   0 \\
                   SMM97C & 18 20.0 34.7 & -16 00 36 &  0.11 &     4.7$\pm$0.6 &        20$\pm$3 &        34$\pm$6 &   0 \\
                   SMM96B & 18 20.0 34.8 & -16 04 34 &  0.13 &     3.9$\pm$0.5 &        15$\pm$3 &        27$\pm$5 &   0 \\
                   SMM95A & 18 20.0 35.1 & -16 08 32 &  0.16 &     6.2$\pm$0.7 &        25$\pm$5 &        44$\pm$9 &   0 \\
                   SMM98A & 18 20.0 35.2 & -16 06 06 &  0.15 &         9$\pm$1 &        34$\pm$7 &       60$\pm$10 &   0 \\
                   SMM96A & 18 20.0 35.2 & -16 04 58 &  0.14 &     4.1$\pm$0.5 &        17$\pm$3 &        29$\pm$6 &   0 \\
                   SMM97A & 18 20.0 35.4 & -16 00 44 &  0.13 &     5.4$\pm$0.6 &        21$\pm$4 &        37$\pm$7 &   1 \\
                   SMM98C & 18 20.0 35.8 & -16 05 32 &  0.12 &     3.3$\pm$0.4 &        13$\pm$2 &        22$\pm$4 &   0 \\
                   SMM97B & 18 20.0 36.7 & -16 00 36 &  0.11 &     3.5$\pm$0.4 &        11$\pm$2 &        19$\pm$3 &   0 \\
                   SMM98B & 18 20.0 36.7 & -16 05 54 &  0.11 &     2.8$\pm$0.3 &         9$\pm$2 &        16$\pm$3 &   0 \\
                  SMM100A & 18 20.0 36.9 & -15 59 46 &  0.17 &     3.2$\pm$0.4 &        20$\pm$4 &        34$\pm$6 &   0 \\
                   SMM99A & 18 20.0 37.0 & -15 58 02 &  0.06 &     2.9$\pm$0.3 &     4.3$\pm$0.7 &         7$\pm$1 &   0 \\
                  SMM101A & 18 20.0 37.6 & -16 01 54 &  0.07 &     4.2$\pm$0.5 &         7$\pm$1 &        11$\pm$2 &   0 \\
                  SMM102A & 18 20.0 37.6 & -15 57 38 &  0.12 &     2.3$\pm$0.3 &         8$\pm$1 &        13$\pm$3 &   0 \\
                  SMM104A & 18 20.0 37.7 & -16 02 52 &  0.06 &     1.7$\pm$0.2 &     2.5$\pm$0.4 &     4.3$\pm$0.6 &   0 \\
                  SMM108A & 18 20.0 38.1 & -15 59 12 &  0.14 &     5.2$\pm$0.6 &        19$\pm$3 &        32$\pm$6 &   0 \\
                  SMM106A & 18 20.0 38.1 & -16 01 36 &  0.14 &     5.2$\pm$0.6 &        17$\pm$3 &        30$\pm$6 &   0 \\
                  SMM109A & 18 20.0 38.3 & -15 58 06 &  0.09 &     3.1$\pm$0.4 &         7$\pm$1 &        11$\pm$2 &   0 \\
                  SMM104B & 18 20.0 38.5 & -16 02 40 &  0.05 &     1.4$\pm$0.2 &     1.9$\pm$0.2 &     3.2$\pm$0.4 &   0 \\
                   SMM95B & 18 20.0 38.7 & -16 08 14 &  0.08 &     1.6$\pm$0.2 &     4.0$\pm$0.6 &         7$\pm$1 &   0 \\
                  SMM112A & 18 20.0 39.4 & -16 05 58 &  0.10 &     2.0$\pm$0.2 &         6$\pm$1 &        11$\pm$2 &   0 \\
                  SMM113A & 18 20.0 39.7 & -16 06 58 &  0.11 &     1.8$\pm$0.2 &         7$\pm$1 &        12$\pm$2 &   0 \\
                  SMM118A & 18 20.0 41.5 & -16 08 28 &  0.07 &     2.3$\pm$0.3 &     4.6$\pm$0.7 &         8$\pm$1 &   0 \\
                  SMM121A & 18 20.0 42.3 & -16 08 40 &  0.04 &     2.1$\pm$0.2 &     2.3$\pm$0.3 &     4.0$\pm$0.5 &   0 \\
                  SMM121B & 18 20.0 43.0 & -16 08 24 &  0.10 &     3.1$\pm$0.4 &         8$\pm$1 &        14$\pm$3 &   1 \\
\enddata
\tablenotetext{a}{The names of the \four\ clumps have been set to reflect the names of the \eight
clumps in which their peaks appear.  Thus, clumps SMM~21A--21C are the 
three 
\four\ clumps whose peaks appear within the boundaries of \eight\ clump 
SMM~21.}
\tablenotetext{b}{The uncertainties stated in this table are composed of
the uncertainties in the gain calibration, the sky opacities, and the
corrections due to the error beam.  These uncertainties are typically
significantly larger than the random errors associated with the rms flux
of the map.  The exception is the peak flux, where the random error of
$\sigma = 0.32$ Jy beam$^{-1}$ dominates the systematic error for the
clumps with lower peak fluxes.}
\tablenotetext{c}{Number of MSX point sources contained within the clump's 0.5$S_{\rm peak}$ contour.}
\end{deluxetable}

\clearpage

\begin{deluxetable}{ccccccccc}
\tabletypesize{\scriptsize}
\tablewidth{0pt}
\tablecaption{Mass Function Parameters of Best Fit\label{tab:m17params}}
\tablehead{
\colhead{} & 
\colhead{Waveband} & 
\colhead{} & 
\colhead{} & 
\colhead{$M_{\rm peak}$\tablenotemark{c}} &
\colhead{$\langle M\rangle$\tablenotemark{c}} &
\colhead{$M_{\rm break}$} &
\colhead{} &
\colhead{} \\
\colhead{Fit Type} & 
\colhead{($\mu$m)} & 
\colhead{$A_{0}$} & 
\colhead{$A_{1}$} & 
\colhead{($M_{\odot}$)} &
\colhead{($M_{\odot}$)} &
\colhead{($M_{\odot}$)} &
\colhead{$\alpha_{\rm low}$} &
\colhead{$\alpha_{\rm high}$} }
\startdata
N($>M$)\tablenotemark{a} & 850 & 2.5 $\pm$ 0.1 & 1.06 $\pm$ 0.07 & 3.9$\pm$ 0.5 & 21$\pm$2 & 17$^{+19}_{-3}$ & -1.35$^{+0.03}_{-0.11}$ & -2.7$^{+0.2}_{-1.0}$ \\
$\Delta N/\Delta M$\tablenotemark{b} & 850 & 2.6 $\pm$ 0.1 & 1.3 $\pm$ 0.1 & 2.5$\pm$0.5 & 32$\pm$3 & 8 $\pm$ 2 & -0.01 $\pm$ 0.26 & -1.5 $\pm$ 0.1 \\
N($>M$)\tablenotemark{a} & 450 & 2.54 $\pm$ 0.07 & 0.87 $\pm$ 0.09 & 5.7$\pm$0.6 & 19$\pm$1 & 18$\pm$4 & -1.5 $\pm$ 0.2 & -3.1 $\pm$ 0.4 \\
$\Delta N/\Delta M$\tablenotemark{b} & 450 & 2.7 $\pm$ 0.1 & 1.2 $\pm$ 0.1 & 3$\pm$1 & 30$\pm$5 & 19$\pm$6 & -0.5$\pm$ 0.2 & -1.9 $\pm$ 0.3 \\
\enddata
\tablenotetext{a}{Uncertainties on fits to the CMFs correspond to the 95\% confidence interval on either side of 
the most likely value of the fitted parameter, as determined from 10$^{5}$ Monte Carlo simulations of the data.  The unequal upper 
and lower uncertainty bounds on the fit to the \eight\ CMF reflect the skew of the distribution of the fitted parameters.}
\tablenotetext{b}{Uncertainties on fits to the DMFs were calculated using a standard nonlinear least-squares fit to the binned data.  
The stated uncertainties are 2~$\sigma$ uncertainties, to match the 95\% 
confidence interval used for the CMF uncertainties.}
\tablenotetext{c}{The mean mass, $\langle M\rangle$, and peak mass, $M_{\rm peak}$ of the lognormal distribution are not parameters of the fit; 
they are calculated from $A_{0}$ and $A_{1}$ using the transformation equations $\langle M\rangle = \exp(A_{0} + \frac{1}{2}A_{1}^{2})$ and 
$M_{\rm peak} = \exp(A_{0} - A_{1}^{2})$.}
\end{deluxetable}


\begin{thebibliography}{}
\bibitem[Adams \& Fatuzzo(1996)]{af96}Adams, F.~C., \& Fatuzzo, M.  1996, \apj, 464, 256
\bibitem[Bonnell et al.(1997)]{bon97}Bonnell, I.~A., Bate, M.~R., Clarke, C.~J., \& Pringle, J.~E.  1997, \mnras, 285, 201
\bibitem[Bonnell et al.(2001a)]{bon01a}Bonnell, I.~A., Bate., M.~R., Clarke, C.~J., \& Pringle, J.~E.  2001a, \mnras, 323, 785
\bibitem[Bonnell et al.(2001b)]{bon01b}Bonnell, I.~A., Clarke, C.~J., Bate, M.~R., Pringle, J.~E.  2001b, \mnras, 324, 573
\bibitem[Bonnell \& Bate(2002)]{bb02}Bonnell, I.~.A., \& Bate, M.~.R.  2002, \mnras, 336, 659
\bibitem[Bonnor(1956)]{b56}Bonnor, W.~G.  1956, \mnras, 116, 351
\bibitem[Brogan \& Troland(2001)]{bt01}Brogan, C.~L., \& Troland, T. H.  2001, \apj, 560, 821
\bibitem[Casassus et al.(2000)]{casassus}Casassus, S., Bronfman, L., May, J., \& Nyman, L.-\AA  2000, \aap, 358, 514
\bibitem[Chabrier(2003)]{chabrier03}Chabrier, G.  2003, \pasp, 115, 763
\bibitem[Chini \& Wargau(1998)]{cw98}Chini, R., \& Wargau, W.~F.  1998, \aap, 329, 161
\bibitem[Chini et al.(2000)]{chini2000l}Chini, R., Nielbock, M., \& Beck, R.  2000, \aap, 357, L33
\bibitem[Dupac et al.(2002)]{dupac02}Dupac, X., et al.  2002, \aap, 392, 691
\bibitem[Ebert(1955)]{eb55}Ebert, R.  1955, Z. Astrophys., 37, 217
\bibitem[Emerson(1995)]{emerson2}Emerson, D.~T.  1995, in ASP Conf. Ser. 75, Multi-Feed Systems for Radio Telescopes, ed. D.~T.~Emerson \& J.~M.~Payne (San Francisco: ASP), 309
\bibitem[Felli et al.(1984)]{fmc84}Felli, M., Massi, M., \& Churchwell, E. 1984, \aap, 136, 53
\bibitem[Hanson et al.(1997)]{hanson97}Hanson, M.~M., Howarth, I.~D., Conti, P.~S.  1997, \apj, 489, 698
\bibitem[Henning et al.(1998)]{henning98}Henning, Th., Klein, R., Launhardt, R., Lemke, D., \& Pfau, W.  1998, \aap, 332, 1035
\bibitem[Hildebrand(1983)]{h83}Hildebrand, R. H.  1983, \qjras, 24, 267
\bibitem[Jaffe et al.(1981)]{jgd81}Jaffe, D.~T., G\"{u}sten, \& Downes, D.  1981, \apj, 250, 621
\bibitem[Johnstone et al.(2000)]{dj2000}Johnstone, D.~J., Wilson, C.~D., Moriarty-Schieven, G., Joncas, G., Smith, G., Gregersen, E., \& Fich, M.  2000, \apj, 545, 327
\bibitem[Johnstone et al.(2001)]{dj2001}Johnstone, D.~J., Fich, M., Mitchell, G.~F., Moriarty-Schieven, G.  2001, \apj, 559, 307
\bibitem[Kerton et al.(2001)]{kerton}Kerton, C.~R., Martin, P.~G., Johnstone, D., \& Ballantyne, D.~R.  2001, \apj, 552, 601
\bibitem[Klessen(2001)]{klessen01}Klessen, R.~S.  2001, \apj, 556, 837
\bibitem[Klessen \& Burkert(2000)]{kb00}Klessen, R., \& Burkert, A.  2000, \apjs, 128, 287
\bibitem[Klein et al.(1999)]{klein99}Klein, R., Henning, Th., \& Cesarsky, D.  1999, \aap, 343, L53
\bibitem[Kramer et al.(1998)]{kramer98}Kramer, C., Stutzki, J., R\"{o}hrig, R., \& Corneliussen, U.  1998, \aap, 329, 249
\bibitem[Kroupa(2002)]{kroupa}Kroupa, P. 2002, Science, 295, 82
\bibitem[Krumholz(2005)]{krum05}Krumholz, M.~R., McKee, C.~F., \& Klein, R.~I.  2005, \apj, 618, 33
\bibitem[Lada(1976)]{l76}Lada, C.~J.  1976, \apjs, 32, 603
\bibitem[Lada et al.(1976)]{lada76}Lada, C.~J., Dickinson, D.~F., Gottlieb, C.~A., \& Wright, E.~L. 1976, \apj, 207, 113
\bibitem[Larson(1973)]{larson73}Larson, R~.B.  1973, \mnras, 161, 133
\bibitem[Ma\'iz Apell\'aniz \& \'Ubeda(2005)]{apellaniz}Ma\'iz Apell\'aniz, J. \& \'Ubeda L. 2005, \apj, 629, 873
\bibitem[McKee \& Tan(2002)]{mt02}McKee, C.~F., \& Tan, J.~C.  2002, \nat, 416, 59
\bibitem[McKee \& Tan(2003)]{mt03}McKee, C.~F., \& Tan, J.~C.  2003, \apj, 585, 850
\bibitem[McLaughlin \& Pudritz(1996)]{mp96}McLaughlin, D.~E. \& Pudritz, R.~E.  1996, \apj, 469, 194
\bibitem[Miller \& Scalo(1979)]{ms79}Miller, G~.E., \& Scalo, J.~M.  1979, \apjs, 41, 513
\bibitem[Mookerjea et al.(2004)]{mookerjea}Mookerjea, B., Kramer, C., Nielbock, M., \& Nyman, L.-\AA. 2004, \aap, 426, 119
\bibitem[Motte et al.(2001)]{m01}Motte, F., Andr\'e, P., Ward-Thompson, D., Bontemps, S.  2001, \aap, 372, L41
\bibitem[Motte et al.(1998)]{man98}Motte, F., Andr\'e, P., \& Neri, R.  1998, \aap, 336, 150
\bibitem[Motte et al.(2003)]{msl03}Motte, F., Schilke, P., \& Lis, D.C.  2003, \apj, 582, 277
\bibitem[Nielbock et al.(2001)]{n01}Nielbock, M., Chini, R., J\"{u}tte, M., \& Manthey, E.  2001, \aap, 377, 273
\bibitem[Press et al.(1992)]{press}Press, W.~H., Teukolsky, S.~A., Vetterling, W.~T., \& Flannery, B.~P.  1992, Numerical Recipes in FORTRAN, (2nd ed.; Cambridge: University Press)
\bibitem[Reid \& Wilson(2005)]{paperi}Reid, M.~A., \& Wilson, C.~D.  2005, \apj, 625, 891 
\bibitem[Sagar \& Richtler(1991)]{sr91}Sagar, R. \& Richtler, T.  1991, \aap, 250, 324
\bibitem[Salpeter(1955)]{sal55}Salpeter, E.~E.  1955, \apj, 121, 161
\bibitem[Sandell \& Sievers(2004)]{ss04}Sandell, G., \& Sievers, A.  2004, \apj, 600, 269
\bibitem[Scalo(1998)]{scalo98}Scalo, J.  1998, in ASP Conf. Ser. 142, The Stellar Initial Mass Function Proceedings, ed. G.~Gilmore \& D. Howell (San Francisco: ASP), 201
\bibitem[Stutzki \& G\"{u}sten(1990)]{sg90}Stutzki, J. \& G\"{u}sten, R.  1990, \apj, 356, 513
\bibitem[Testi \& Sargent(1998)]{ts98}Testi, L. \& Sargent, A.~I.  1998, \apj, 508, L91
\bibitem[Tothill et al.(2002)]{tot}Tothill, N.~F.~H., White, G.~J., Matthews, H.~E., McCutcheon, W.~H., McCaughrean, M.~J., Kenworthy, M.~A. 2002, \apj, 580, 285
\bibitem[Williams et al.(1994)]{clfind}Williams, J.~P., de~Geus, E.~J., \& Blitz, L.  1994, \apj, 428, 693
\bibitem[Wilson et al.(1999)]{wilson99}Wilson, C.~D., Howe, J.~E., \& Balogh, M.~L. 1999, \apj, 517, 174
\bibitem[Wilson et al.(1979)]{wilson79}Wilson, T.~L., Fazio, G.~G., Jaffe, D., Kleinmann, D., Wright, E.~L., \& Low, F.~J. 1979, \aap, 76, 86
\bibitem[Young et al.(2004)]{young04}Young, C.~H., et al.  2004, \apjs, 154, 396 
\bibitem[Zinnecker(1984)]{z84}Zinnecker, H.  1984, \mnras, 210, 43
\end{thebibliography}
\end{document}